%% file: paper.tex
\documentclass[floatfix,aps,pra,superscriptaddress, longbibliography,twocolumn]{revtex4-2}
\usepackage{color}
\usepackage{xcolor}
\usepackage{amsmath}
\usepackage{amsthm}
\usepackage{amssymb}
\usepackage{graphicx}
\usepackage{caption}
\usepackage{booktabs}
\usepackage{ragged2e}
\usepackage{float}
\usepackage{hyperref}

\begin{document}
\include{parts/Main}

\include{parts/Supplement}

\end{document}

%% file: parts/Main.tex
\captionsetup{
    margin=5pt,                
    font=small,                
    justification=Justified,   
    singlelinecheck=false,
    position=below
}
\setlength{\parindent}{0pt}


\title{\LARGE Integrating experimental feedback improves generative models for biological sequences}
\author{Francesco Calvanese}
\thanks{These authors contributed equally to this work.}
\affiliation{Sorbonne Universit\'{e}, CNRS, Dept.~of Computational, Quantitative and Synthetic Biology - CQSB,  Paris, France}
\affiliation{Laboratoire de Biophysique et Evolution, CNRS-ESPCI, PSL University, Paris, France}

\author{Giovanni Peinetti}
\thanks{These authors contributed equally to this work.}
\affiliation{Sorbonne Universit\'{e}, CNRS, Dept.~of Computational, Quantitative and Synthetic Biology - CQSB,  Paris, France}

\author{Polina Pavlinova}
\affiliation{Laboratoire de Biophysique et Evolution, CNRS-ESPCI, PSL University, Paris, France}

\author{Philippe Nghe}
\affiliation{Laboratoire de Biophysique et Evolution, CNRS-ESPCI, PSL University, Paris, France}

\author{Martin Weigt}
\email{To whom correspondence should be addressed. \\E-mail: martin.weigt@sorbonne-universite.fr}
\affiliation{Sorbonne Universit\'{e}, CNRS, Dept.~of Computational, Quantitative and Synthetic Biology - CQSB,  Paris, France}

\begin{abstract}
\hspace{-3.5mm}\textbf{Generative probabilistic models have shown promise in designing artificial RNA and protein sequences but often suffer from high rates of false positives, where sequences predicted as functional fail experimental validation. To address this critical limitation, we explore the impact of reintegrating experimental feedback into the model design process. We propose a likelihood-based reintegration scheme, which we test through extensive computational experiments on both RNA and protein datasets, as well as through wet-lab experiments on the self-splicing ribozyme from the group I intron RNA family where our approach demonstrates particular efficacy. We show that integrating recent experimental data enhances the model's capacity of generating functional sequences (e.g. from 6.7\% to 63.7\% of active designs at 45 mutations). This feedback-driven approach thus provides a significant improvement in the design of biomolecular sequences by directly tackling the false-positive challenge.}
\end{abstract}

\maketitle

\section{Introduction}

Generative probabilistic models for biological sequences, such as proteins and RNA, have recently emerged as promising tools for designing artificial biomolecules \cite{russetal, Lambert2024-rd, LLM, remisimona}. These models, particularly family-specific ones like those built using Direct-Coupling Analysis (DCA) \cite{russetal, Lambert2024-rd}, as well as more advanced architectures like restricted Boltzmann machines \cite{remisimona}, variational autoencoders \cite{vae1, vae2, Lambert2024-rd}, and protein language models \cite{LLM}, have shown notable success in generating functional sequences. However, a persistent challenge remains: these models often produce a high rate of false positives -- sequences generated as potentially functional by the model but failing in experimental tests.

\enlargethispage{-20.1pt}

These models are trained on sets of homologous sequences, representing families of sequences with shared evolutionary ancestry. Such families are typically characterized by highly conserved structures and functions, though the sequences themselves may diverge significantly. Multiple-sequence alignments (MSA) \cite{Rfam, Pfam, Infernal,Hmmer}, containing presumably functional sequences from different species, serve as the foundation for training. As a consequence, these models are trained in an unsupervised manner on unlabeled functional sequences, which limits their capacity to differentiate between functional and non-functional variants.

A significant issue in these generative models is the high rate of false positives -- sequences deemed functional by the model that fail experimental validation \cite{russetal, Lambert2024-rd}. This limitation arises from the scarce sampling of the viable sequence space in the MSAs, leading to an underrepresentation of functional diversity, and to an intrinsic difficulty in accurately estimating the limitations of functional sequence space.

In this study, we focus on DCA-based Potts models \cite{russetal, muntoni2021adabmdca, cuturello2020assessing} and demonstrate that integrating experimental feedback including false-positive sequences into the training procedure can enhance model accuracy and reduce false-positive rates. By incorporating this feedback through an extension of the maximum-likelihood inference procedure \cite{cuturello2020assessing}, which makes explicit use of the experimental results, we show that false positives from the initial model play a critical role in refining the boundaries of the viable sequence space, thereby improving the model's performance.
An intriguing ingredient to this approach is that the underlying mathematical structure of the model remains unchanged, but the reintegration of experimental data significantly improves parameter learning. This highlights an important insight: the current limitations of generative models stem not necessarily from the limited expressivity of their architectures, but also from the insufficient information content in the original training data. These alignments, representing natural sequences that have diverged through evolution, offer a sparse and incomplete sampling of the functional sequence landscape. Enhancing this landscape with experimental feedback allows for a reliable model, generating a higher fraction of functional sequences (i.e.~true positives). Augmenting data quantity and quality at unchanged model complexity turns out to be an efficient strategy.

The paper is organized as follows. In the next \emph{Results} section, we outline the main ideas of the reintegration approach and present extensive tests on diverse RNA and protein families, progressing from purely computational settings to experimental validation : artificial sequences sampled from our models and tested experimentally are the most rigorous possible validation of a generative approach to bio-molecules. We detail the application of our procedure to DCA models in the \emph{Materials and Methods} section, where we also describe the datasets used to evaluate our approach. At the end of the article, we present our main conclusion, and show an outlook of possible extensions of our approach.

\begin{figure*}[t]
\begin{center}
\includegraphics[width=1\textwidth]{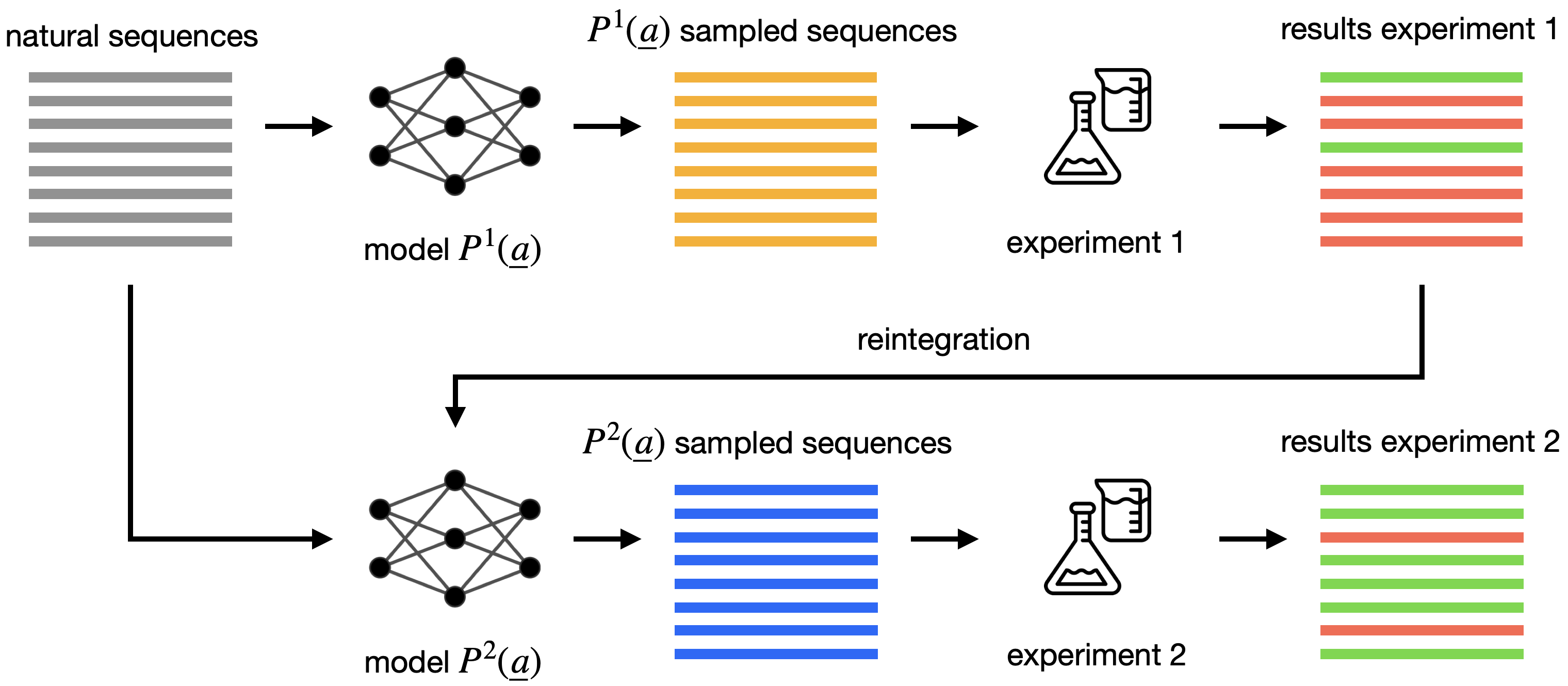}
\end{center}
\caption{ \textit{Schematic representation of the reintegration procedure. The original model probability distribution \( P^1(\underline{a}) \) trained on a natural MSA is used to sample the \(P^1\)-dataset. These sequences are experimentally tested and labeled, and this information is reintegrated into the training of a DCA model \( P^2(\underline{a}) \). The new \(P^2\)-dataset exhibits enhanced functionality. Note that the mathematical form of $P^1(\underline{a})$ and $P^2(\underline{a})$ is identical, and that the improved generative performance results from refined parameter values learned on enriched data.}}
\label{reintegration}
\end{figure*}

\section{RESULTS}

Here we propose a method to reintegrate experimental test results into a sequence generative model.
Figure~\ref{reintegration} illustrates the core idea. In the standard approach, the natural data MSA \(\mathcal{D}_N\) is used to train an initial probabilistic generative model \(P(\underline{a} \, | \, \theta^1 ) = P^1(\underline{a} )\) whose parameters \(\theta^1\) are obtained through Maximum Likelihood Estimation (MLE):
\begin{equation}
\theta^1 = \arg\max_{\theta} \; \mathcal{L}(\theta \, | \, \mathcal{D}_N),
\label{eq:MLE_theta1}
\end{equation}
where
\begin{equation}
\mathcal{L}(\theta \, | \, \mathcal{D}_N)
= \frac{1}{|\mathcal{D}_N|} \sum_{\underline{a} \in \mathcal{D}_N} \ln P(\underline{a} \, | \, \theta).
\label{eq:MLE_L}
\end{equation}
Once the model is trained, an set \(\mathcal{D}_T\) of artificial sequences can be sampled from \(P^1(\underline{a} )\) and tested experimentally. However, this approach often suffers from a high rate of false-positive sequences in \(\mathcal{D}_T\), i.e.\ sequences expected to be functional according to $P^1(\underline{a})$, yet failing experimental tests (indicated in red in Figure~\ref{reintegration}), cf.~\cite{russetal,Lambert2024-rd}. To address this issue, we propose reintegrating the experimental feedback contained in \(\mathcal{D}_T\) into an updated model, \(P(\underline{a} \, | \, \theta^2) = P^2(\underline{a}) \). This updated model maintains the same mathematical form and architecture as \(P^1(\underline{a} )\) but uses recalibrated parameters inferred leveraging the newly labeled data \(\mathcal{D}_T\). Consequently, \(P^2(\underline{a})\) is expected to generate a higher proportion of functional (true-positive) sequences. To implement this, we update the model parameters optimizing a new objective function:
\begin{equation}
\theta^2 = \arg\max_{\theta} \; \mathcal{Q}(\theta \, | \, \mathcal{D}_N, \mathcal{D}_T ),
\label{eq:MLE_theta1}
\end{equation}
where
\begin{eqnarray} \label{eq:new_objective}
\mathcal{Q}( \theta \, | \, \mathcal{D}_N, \mathcal{D}_T) &=&
\mathcal{L}(  \theta \, | \, \mathcal{D}_N) + \\
&&+ \frac{\lambda}{|\mathcal{D}_T|}\sum_{\underline{b} \in  \mathcal{D}_T} w(\underline{b}) \cdot \ln P(\underline{b} \, | \,  \theta)\ ,
\nonumber 
\end{eqnarray}
The first contribution to \(\mathcal{Q}\) equals the standard log-likelihood for the natural data given in Eq.~\eqref{eq:MLE_L}. Maximizing \(\mathcal{Q}\) reduces thus to standard MLE if no experimental data is available, i.e., for \( \mathcal{D}_T = \emptyset\).
The second contribution is the reintegration term, which acts on the probabilities of the tested sequences in dependence on the adjustment weight \(w\), assigned to every sequence in the tested dataset \(\underline{b} \in  \mathcal{D}_T\) in function of the experimental test result. We require \(w(\underline{b})\) to adhere to the following rules:
\begin{itemize}
    \item Negative Adjustment: \(w(\underline{b}) < 0\) for sequences failing the experimental functionality test, such that their probability \(P^2(\underline{b})\) is reduced when maximizing \(\mathcal{Q}\).
    \item Positive Adjustment: \(w(\underline{b}) > 0\) for sequences passing the experimental functionality test, such that their probability \(P^2(\underline{b})\) is increased when maximizing \(\mathcal{Q}\).
\end{itemize}
The specific values of the weights for individual sequences depend on the specific experimental setting, in the easiest case they can be taken to be all of equal absolute value (see below for some more complicated construction removing at least partially biases in the experimental data). The overall intensity of reintegration is controlled by the hyperparameter \(\lambda\); the higher is \(\lambda\), the greater the relative importance assigned to the experimentally labelled dataset \(\mathcal{D}_T\) compared to the natural sequences \(\mathcal{D}_N\). When \(\lambda = 0\), the classical MLE is recovered. If we consider only the functional sequences with \(w(\underline{b}) > 0\) in \( \mathcal{D}_T\), this procedure is similar to adding them to \( \mathcal{D}_N\) with a $\lambda$-dependent weight, and performing the standard MLE inference. 

\begin{figure}[htbp]
\begin{center}
\includegraphics[width=\columnwidth]{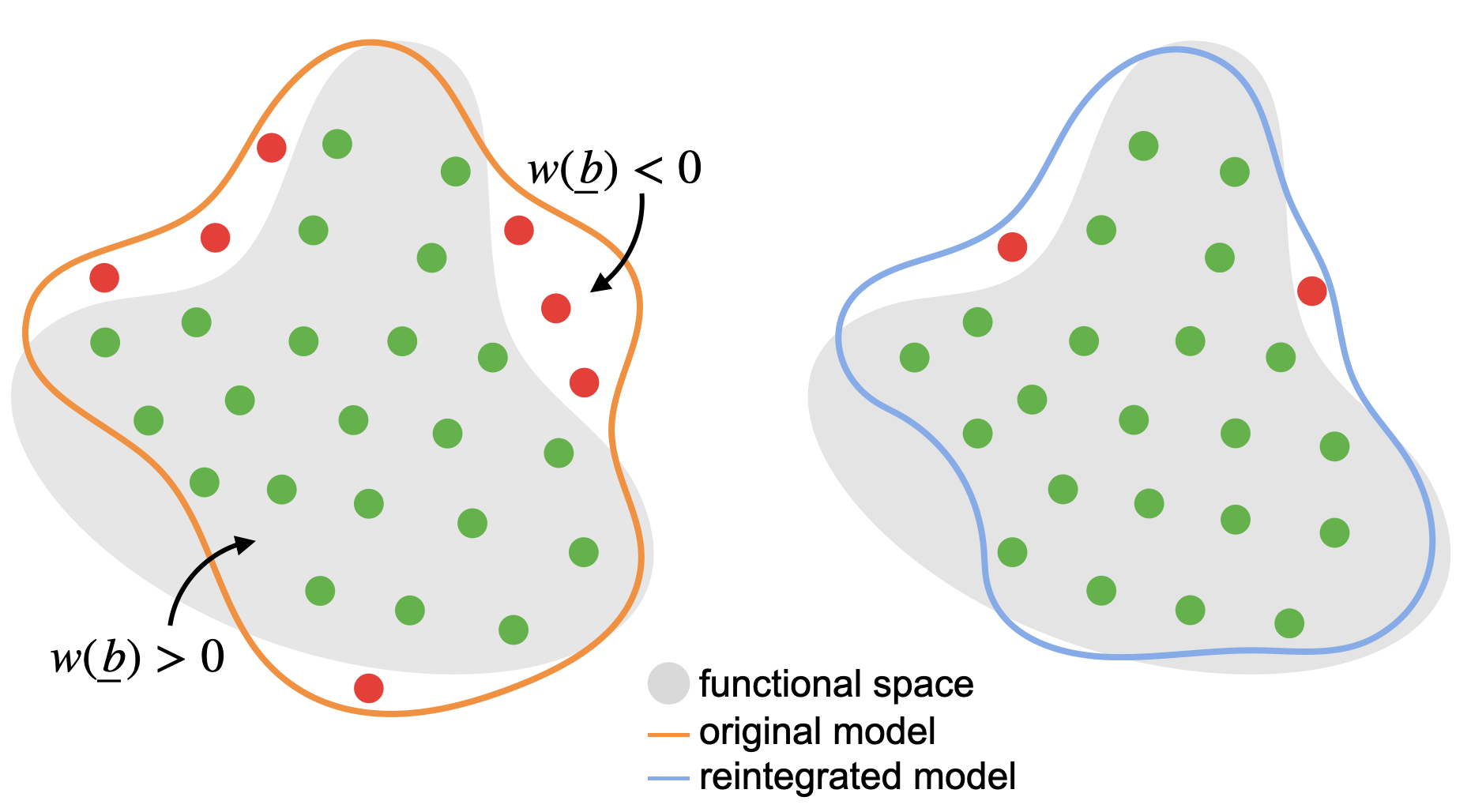}
\end{center}
\caption{\textit{Stylized representation of the effect of the reintegration procedure. Sequences generated by the initial model \( P^1 \) (region inside orange line) that fail experimental tests (region outside grey area) are assigned negative adjustment weights (\( w < 0 \)), while those that are functional are assigned positive weights (\( w > 0 \)). The reintegrated model \( P^2 \) is then trained using these adjusted weights, resulting in generated sequences that avoid regions associated with non-functional sequences and concentrate in regions associated with functional sequences (region inside blue line), thereby reducing the fraction  of false positives among the generated sequences.}}
\label{procedure}
\end{figure}

The essential difference arises from the inclusion of non-functional sequences with \(w(\underline{b}) < 0\) in \( \mathcal{D}_T\), which indicate regions of the sequence space that our model should avoid, cf.~Fig.~\ref{procedure}.
Ideally, \( \mathcal{D}_T\) would consist of $P^1(\underline{a} )$ generated sequences, cf.~Fig.~\ref{reintegration}. Consider a sequence \(\underline{b}\) that has been generated by \(P^1(\underline{a} )\): the sole fact that it was sampled implies that it was assigned a high probability by the \(P^1(\underline{a} )\) model. If this sequence fails the experimental test, it is assigned a negative \(w(\underline{b}) < 0\), and the reintegrated \(P^2(\underline{a})\) model will subsequently assign it a lower probability. This procedure enables the model to correct itself based on the experimental feedback, and to better infer the limits of functional sequence space.

The reintegration set \( \mathcal{D}_T\) is, however, not limited to be comprised of sequences generated by \(P^1(\underline{a}  )\) but can, in principle, be any functionally labelled dataset. It is, however, intuitively important that negative sequences are close to the functional sequence space. Random sequences, e.g., are almost surely non-functional, but they typically have already very low values of $P^1(\underline{a} )$, and reintegrating them negatively will not improve the description of the positively functional sequence space. 

Our tests of this procedure have been carried out on DCA models, since DCA has proved capable of generating functional proteins and RNAs \cite{russetal, Lambert2024-rd}. Another advantage is that we can naturally implement the new objective function \(\mathcal{Q}\) in the DCA framework without altering its classical training procedures. A detailed discussion is provided in the \emph{Materials and Methods} section with all analytical derivations detailed in the \emph{Supplementary Section~S1}.
Note that, in a more generic machine-learning context, the DCA model can be replaced by other generative model architectures, and MLE by the optimization of any loss function, which is additive in data-point specific losses. 

In the following, we present computational tests on RNA and protein data, as well as experimental validation performed on group~I intron ribozymes.

\subsection{The effect of the reintegration strength $\lambda$ in Rfam RNA families}

To understand the action of the proposed reintegration method, we need to study its performance systematically in function of the reintegration strength $\lambda$. For this aim, we use three RNA-family MSAs from the Rfam database (cf.\ \emph{Materials and Methods}). Before performing resource and time consuming experiments (cf.~below), we first assess the role of $lambda$ via a fully computational approach. As a proxy for the experimental fitness, we employ the negative free energy \(-F\) of folding a given sequence onto the family’s consensus secondary structure, computed from the Turner Model \cite{turner}) implemented in the \texttt{RNAeval} function of the ViennaRNA package~\cite{ViennaRNA}. Further details on the data and fitness proxy are provided in the \emph{Materials and Methods}.

For the statistical models, we use our recent time-efficient RNA-tailored Edge Activation DCA (eaDCA) \cite{10.1093/nar/eaDCA}. It provides easy access to the inferred models' Shannon entropy \( S \) \cite{entropy} in function of $\lambda$, and thereby allows to quantify the potential diversity of the sequences generated by the model.

To this end, for each of the three RNA families, we used the Rfam MSA as \(\mathcal{D}_N\) and trained our initial model \(P^1(\underline{a})\) via eaDCA. From this model, we sampled the $P^1(\underline{a})$-dataset comprising 2000 sequences, to be used as $\mathcal{D}_T$ in the reintegration procedure. We measured the \texttt{RNAeval} proxy fitness $-F$ \cite{ViennaRNA} of these sequences (cf.~\emph{Materials and Methods}), and decided (somewhat arbitrarily) to consider all sequences with above-average $-F$ as functional, and below-average $-F$ as non-functional - the training objective for $P^2(\underline a)$ thus being the generation of highly thermo-stable sequences. We thus define a simple adjustment weight $w(\underline{b})$ for all $\underline{b}\in\mathcal{D}_T$:
\begin{equation}
\label{eq:w_Favg}
    w(\underline{b}) = \left\{
    \begin{array}{lcl}
        +1 &&  \text{if } -F(\underline{b}) \geq -F_{avg}\\
        -1 && \text{if } -F(\underline{b}) < -F_{avg}
    \end{array}
    \right.\ ,
\end{equation}
where $F_{avg}$ is the average \texttt{RNAeval} folding free energy evaluated for the $\mathcal{D}_T$ dataset.

A reasonable range for the reintegration strength $\lambda$ can be chosen using the following consideration: as already mentioned, the case $\lambda=0$ is equivalent to training the standard DCA model $P^1(\underline a)$ using MLE and the natural MSA $\mathcal{D}_N$. For $\lambda=1$, the two contributions to the objective function $\mathcal{Q}$ in Eq.~(\ref{eq:new_objective}) become equally important, and so do the two datasets $\mathcal{D}_N$ and $\mathcal{D}_T$. It is therefore reasonable to explore $\lambda$s between zero and values slightly larger than one. Note that, for larger values, the learning algorithm starts to have convergence problems (see \emph{Materials and Methods}).

To assess the effect of the reintegration procedure, we monitor the following quantities:
\begin{enumerate}
    \item \textit{Average Proxy Fitness:} \(-\overline{F}\) is the average proxy fitness of the sequences in the \( P^2 \)-dataset. This measures how well our reintegration is pushing the generation towards "functional" sequences according to Eq.~(\ref{eq:w_Favg}).
    \item \textit{True Positives}: TP is defined as the fraction of sequences in the \(P^2\)-dataset that exhibit a fitness score \(F(\underline{b}) \geq -F_{avg}\). In other words, these sequences would have been assigned a positive weight during the reintegration procedure, indicating predicted functionality. This metric quantifies the effectiveness of the approach in reducing false positives.

    \item \textit{Model entropy:} \( S \) quantifies the diversity of sequences generated by the reintegrated model $P^2$. A higher entropy suggests a more diverse sequence space.
    \item \textit{Average intra-dataset distance:} \( D_{P^2-P^2} \) is the average Hamming distance (number of mutations) between pairs of sequences in the \(P^2\)-dataset. It quantifies the diversity between generated sequences. This value is to be compared with \( D_{P^1-P^1} \), i.e., the average distance in the non reintegrated \( P^1(\underline{a}) \)-dataset.
    \item \textit{Average minimum distance to functional sequences in $\mathcal{D}_T$:} for each sequence in the \( P^2 \)-dataset, we calculate the Hamming distance to the closest functional sequence in the reintegration dataset \( \mathcal{D}_T \). The average of these is reported as \( D_{P^2-\mathcal{D}_T^+} \). This metric ensures that our model is not merely replicating functional sequences from \( \mathcal{D}_T \).
    \item \textit{Average minimum distance to non-functional sequences in $\mathcal{D}_T$:} for each sequence in the \( P^2 \)-dataset, we calculate the Hamming distance to the closest non-functional sequence in the reintegration dataset \( \mathcal{D}_T \). The average of these is reported as \( D_{P^2-\mathcal{D}_T^-} \). This metric allows us to test if generated sequence avoid the vicinity of non-functional sequences from $\mathcal{D}_T$, as is to be expected by their negative contribution to the objective $\mathcal{Q}$.
\end{enumerate}

By tracking these quantities, we can evaluate whether the reintegration leads to a better model without overfitting the reintegrated data. The results of these analyses are presented in Table~\ref{table:01} for the RF00162 RNA family, similar results are observed also for the other families, cf. ~\emph{Supplementary Section~S2} in ~\emph{Supplementary Tables~S1, S2, S3 } and ~\emph{Supplementary Figure~S1}.
\begin{table}
\begin{tabular*}{\columnwidth}{@{\extracolsep{\fill}}lcccccc@{}}
\toprule
$\lambda$ & $-\overline{F}$ & TP & $S$ & $D_{P^2-P^2}$  & $D_{P^2-\mathcal{D}_T^+}$ & $D_{P^2-\mathcal{D}_T^-}$  \\
\midrule
$0$   & 22.4  & $49.9\%$  & 63.4  & 46.0 & --  & --   \\
$0.1$ & 24.3  & $60.4\%$  & 65.3  & 45.8 & 28.1 & 29.6 \\
$0.5$ & 30.2  & $91.3\%$  & 63.2  & 42.2 & 26.7 & 30.5 \\
$1.0$ & 34.9  & $99.4\%$  & 53.3  & 39.7 & 25.3 & 31.8 \\
\bottomrule
\end{tabular*}

\caption{\textit{Effect of the reintegration strength for the RF00162 RNA family. Note that the values reported for $\lambda=0$ correspond to the DCA model without reintegration, i.e. $-F_{avg}=14.68$ is the threshold value chosen for functional sequences, and the entropy $S=61.92$ equals the one of $P^1$.}}
\label{table:01}%
\end{table}

In general, we observe that higher values of \( \lambda \) yield a higher average proxy fitness \( -\overline{F} \), demonstrating that the reintegration procedure effectively enhances the fitness of the generated sequences. For example, for the RF00162 RNA family (Table \ref{table:01}), the average proxy fitness increases from 22.4 (\( \lambda = 0 \)) to 34.9 (\( \lambda = 1 \)), and the TP fraction improves from 49.9\% to 99.4\%. However, this gain comes at the expense of model entropy \( S \) and overall sequence diversity: \( S \) decreases from 63.4 (\( \lambda = 0 \)) to 53.3 (\( \lambda = 1 \)), while the average intra-dataset distance \( D_{P^2-P^2} \) drops from 46.0 (\( \lambda = 0 \)) to 39.7 (\( \lambda = 1 \)). Additionally, generated sequences become somewhat closer to the functional sequences in the reintegration dataset \( \mathcal{D}_T^+ \), as indicated by a reduction in the average minimum distance \( D_{P^2-\mathcal{D}_T^+} \) from 28.1 (\( \lambda = 0.1 \)) to 25.3 (\( \lambda = 1 \)). The average minimum distance to non-functional sequences \( D_{P^2-\mathcal{D}_T^-} \) slightly increases from 29.6 (\( \lambda = 0.1 \)) to 31.8 (\( \lambda = 1 \)).

\begin{figure}[t]
\begin{center}
\includegraphics[width=\columnwidth]{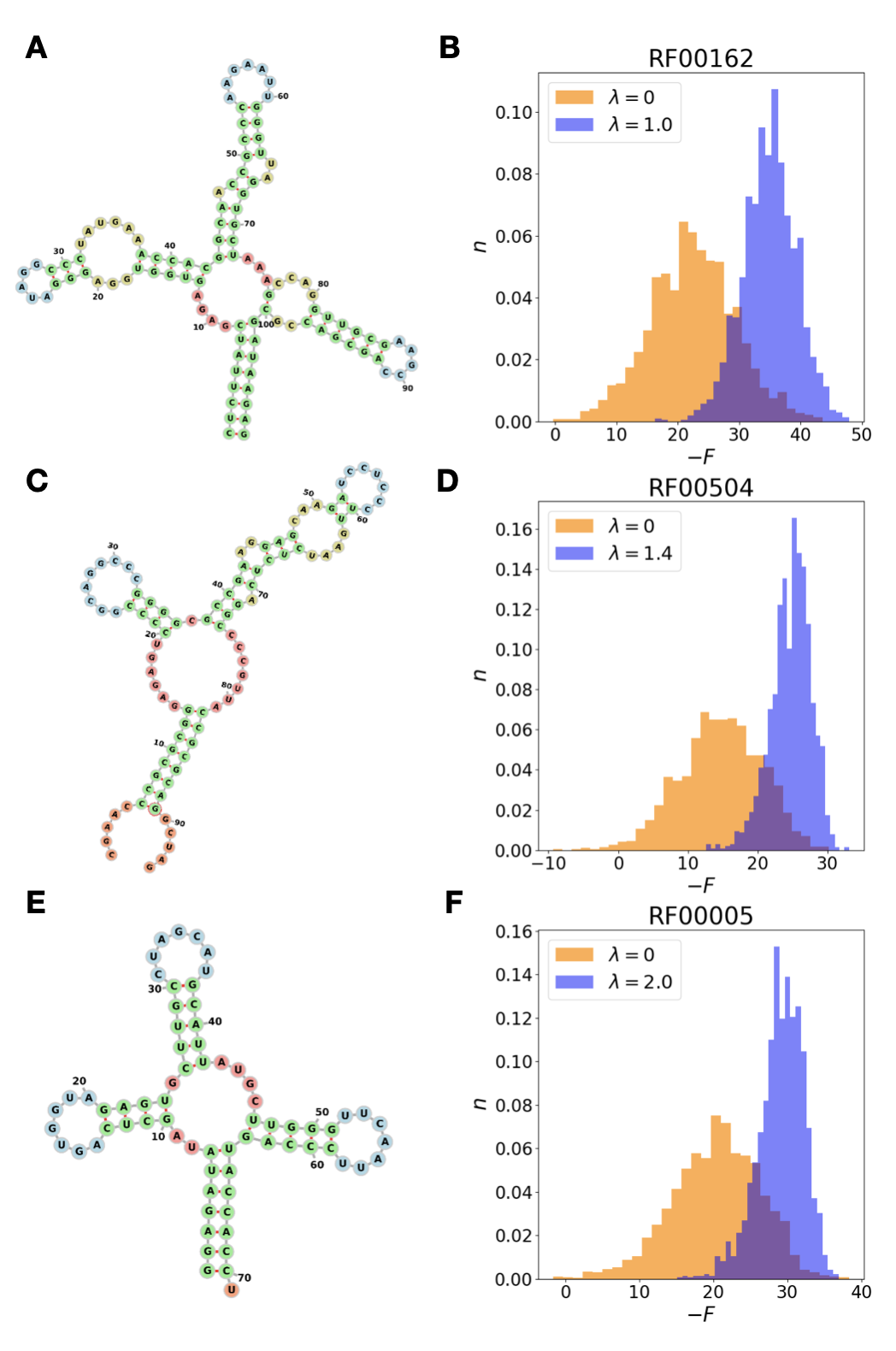}
\end{center}
\caption{\textit{
\textbf{A} Consensus secondary structure of the RF00162 RNA family (SAM riboswitch). 
\textbf{B} Distribution of the \texttt{RNAeval} proxy fitness for the RF00162 DCA model \( P^1(\underline{a}) \) (orange, \( \lambda = 0 \)) and the reintegrated \( P^2(\underline{a}) \) model (blue, \( \lambda = 1 \)).
\textbf{C} Consensus secondary structure of the RF00504 RNA family (Glycine riboswitch). 
\textbf{D} Distribution of the \texttt{RNAeval} proxy fitness for the RF00504 DCA model \( P^1(\underline{a}) \) (orange, \( \lambda = 0 \), $N=2000$) and the reintegrated \( P^2(\underline{a}) \) model (blue, \( \lambda = 2 \), $N=2000$). 
\textbf{E} Consensus secondary structure of the RF00005 RNA family (tRNA). 
\textbf{F} Distribution of the \texttt{RNAeval} proxy fitness for the RF00005 DCA model \( P^1(\underline{a}) \) (orange, \( \lambda = 0 \), $N=2000$) and the reintegrated \( P^2(\underline{a}) \) model (blue, \( \lambda = 1 \), $N=2000$). \\
The secondary structure diagrams were generated using the Forna software \cite{Forna}.
}}
\label{RNAeffect}
\end{figure}

We were able to perform these tests because we can readily compute the proxy fitness also on the sequences generated from the reintegrated model. In more realistic scenarios, this is not possible without experiments. However, our observations guide the choice of $\lambda$ before doing experiments: it seems reasonable to select a value before a significant loss in diversity occurs or before the training procedure fails to converge. 

The effect of reintegration on the proxy fitness \( -F \) distributions for the three Rfam families is shown in Fig.~\ref{RNAeffect}. In all cases, the \( P^2 \)-dataset samples exhibit a significant shift to higher proxy fitness compared to the \( P^1 \)-dataset (indicated by \( \lambda = 0 \)). We find in particular, that "non-functional" sequences with proxy fitnesses below $-F_{avg}$ become very rare.

In addition to generating sequences with improved proxy fitness, another effect of the reintegration is that the DCA model score becomes a more reliable predictor of fitness, an in-depth analysis of this phenomenon is provided in the \emph{Supplementary Section~S2} and \emph{Supplementary Figure~S2}

\subsection{Reintegrating experimental activity of a protein family}

Our case study for applying the reintegration procedure to proteins is the chorismate mutase (CM) enzyme, which plays an essential role in the biosynthesis of aromatic amino acids. This enzyme serves as an ideal setting to test our procedure because Russ et al. \cite{russetal} have already trained a DCA model $P^1$ on an MSA $\mathcal{D}_N$ of natural CM homologs, and they have experimentally tested the natural sequences of $\mathcal{D}_N$ as well as a dataset $\mathcal{D}_T$ of $P^1$-designed CM variants using an {\em in vivo} growth assay (see \emph{Materials and Methods}). As a result, we have access to experimentally labeled datasets indicating sequence functionality. Moreover, Russ et al. demonstrated that it is possible to train a simple Logistic Regression (LR) classifier on $\mathcal{D}_N$ to predict whether an artificial CM variant is functional or not. We can leverage these findings for our reintegration procedure.

Our approach begins by training an LR classifier using the labeled natural CM variants in $\mathcal{D}_N$ to predict experimental functionality. This classifier achieves an accuracy of approximately 80\% in predicting the functionality of artificially generated sequences (cf.~\emph{Supplementary Section~S3}), which is consistent with the results reported \cite{russetal}. Since experimentally testing our artificial sequences is not feasible within this study, we will use this classifier to evaluate the performance of our reintegration procedure. Note that the labels used for training the classifier are not used in our reintegration procedure, but are complementary information exploitable for posterior sequence evaluation.

We first trained our initial DCA model $P^1(\underline{a})$ using the natural MSA $\mathcal{D}_N$ as training data, and the adabmDCA implementation of DCA \cite{muntoni2021adabmdca}. From this model, we sampled the $P^1$-dataset comprising 8000 artificial CM variants.

To train our reintegrated DCA model $P^2(\underline{a})$, we used dataset $\mathcal{D}_T$, which is experimentally labelled in \cite{russetal}, allowing us to avoid relying on proxy fitness measures. We chose again a binary adjustment function $w(\underline{b})$ for all sequences in $\mathcal{D}_T$:
\begin{equation}
\label{eq:w_CM}
    w(\underline{b}) = \left\{
    \begin{array}{lcl}
        +1 &&  \text{if }\ \underline{b}\ \text{ is functional }\\
        -1 && \text{if }\ \underline{b}\ \text{ is non-functional }
    \end{array}
    \right.\ .
\end{equation}
We set the reintegration strength parameter $\lambda$ to 1, which is the highest value that converged within an acceptable time frame.
The reintegration procedure is significantly slower for proteins compared to RNA, requiring 3 hours of runtime on an L4 GPU using the most advanced DCA GPU implementation available. Also here, we sampled from the resulting model a $P^2$-dataset containing 8000 artificial CM sequences.
Results for lower values of $\lambda$ are provided in the \emph{Supplementary Table~S4}.

To assess the effectiveness of our reintegration procedure, we employed the LR classifier, and determined the percentage of variants predicted to be functional in both the $P^1$- and the reintegrated $P^2$-dataset.

\begin{figure}[t]
\begin{center}
\includegraphics[width=\columnwidth]{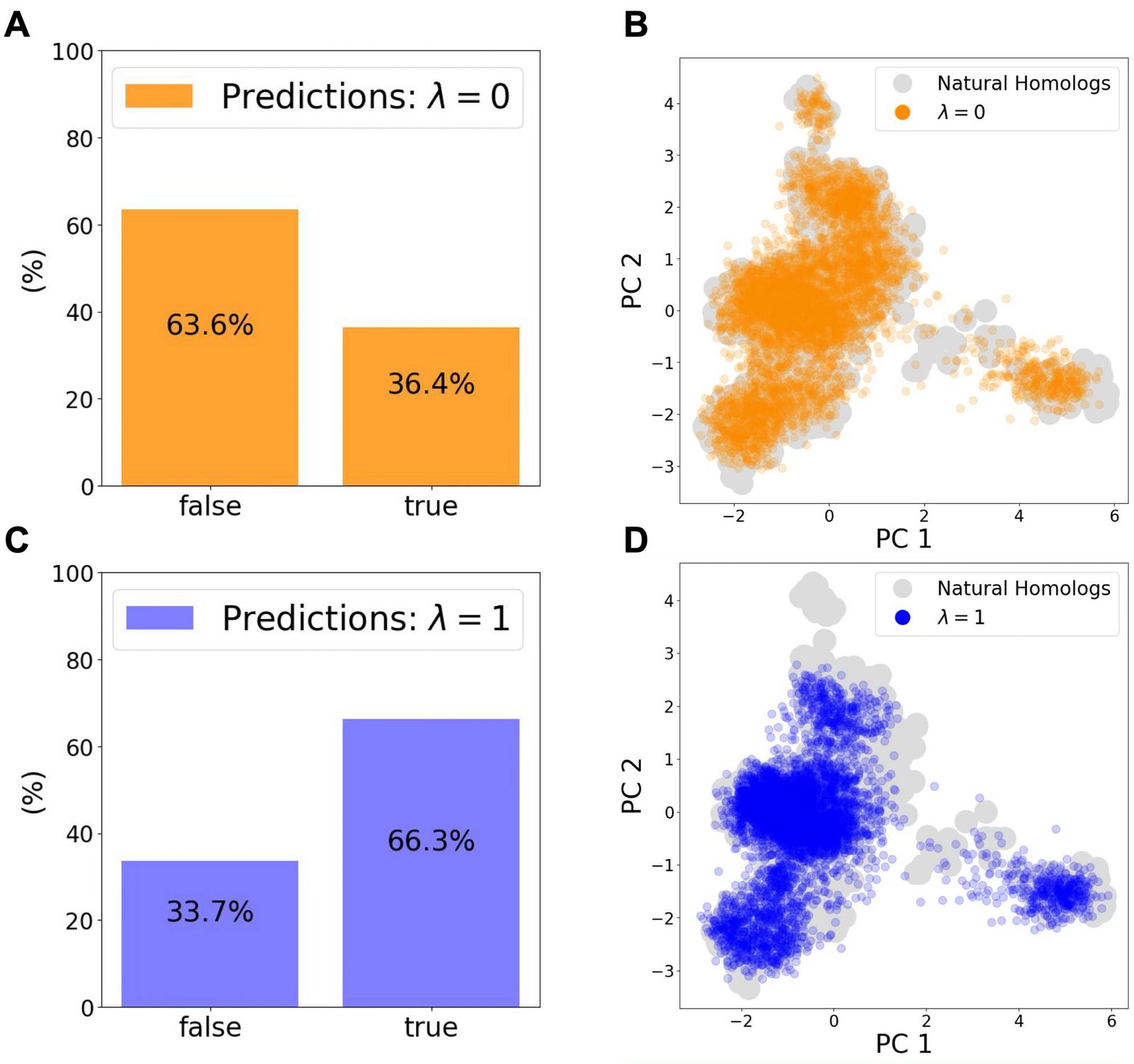}
\end{center}
\caption{\textit{
\textbf{A} Classifier predictions for protein functionality in the \( P^1(\underline{a}) \)-dataset ($N = 8000$): \textit{True} indicates predicted functional, \textit{False} indicates predicted non-functional. \\
\textbf{B} PCA projection of the \( P^1(\underline{a}) \)-dataset (orange); natural PSA homologs are represented by the grey cloud. \\
\textbf{C} Classifier predictions for protein functionality in the \( P^2(\underline{a}) \)-dataset ($N = 8000$): \textit{True} indicates predicted functional, \textit{False} indicates predicted non-functional. \\
\textbf{D} PCA projection of the \( P^2(\underline{a}) \)-dataset (blue); natural PSA homologs are represented by the grey cloud.}}
\label{protein_effect}
\end{figure}
The percentage of predicted functional variants increases from 39\% to 68\%, indicating a favorable outcome of the procedure.  This comes at a moderate cost of reduced average sequence diversity, \( D_{P^1-P^1} - D_{P^2-P^2} = 4.0  \) (5.5\% of \( D_{P^1-P^1} = 72.9 \)), and the designs still retain a good level of diversity (SI). \( D_{P^2-\mathcal{D}_T} \) values and the histogram of distances between the \( P^2 \) dataset and the closest reintegrated \( \mathcal{D}_T \) are provided in the SI.
\begin{figure}[t]
\begin{center}
\hspace{0cm}
\includegraphics[width=0.85\columnwidth]{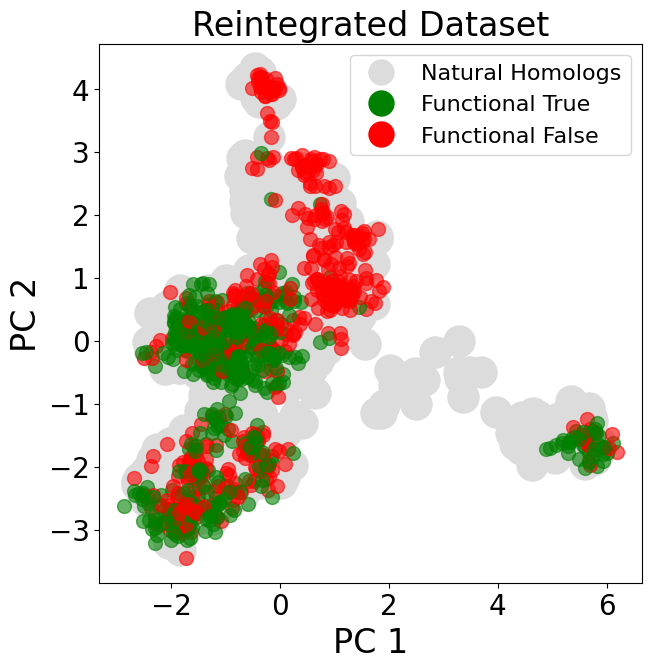}
\end{center}
\caption{\textit{PCA projection of the experimentally labeled dataset $\mathcal{D}_T$ ($N = 1003$) of artificially generated chorismate mutase sequences. Functional sequences are shown in green, non-functional ones in red. The natural sequences in $\mathcal{D}_N$ ($N = 1130$) are shown in grey (note that not all of them are functional in the experimental test performed in \emph{E. coli}).}}
\label{PSB}
\end{figure}

In the PCA plots shown in Fig.~\ref{protein_effect}, we project the $P^1$- and $P^2$-generated datasets (colored) onto the first two principal components of the natural MSA $\mathcal{D}_N$ (grey). Notably, the $P^2$ dataset avoids certain regions in PCA space that are occupied by sequences from both the natural dataset and the $P^1$ model. To better understand this behavior, Fig.~\ref{PSB} projects the reintegration dataset $\mathcal{D}_T$ onto the same PCA, with functional sequences shown in green and non-functional ones in red. We observe that the regions avoided by the $P^2$ dataset correspond to the non-functional areas in $\mathcal{D}_T$. These findings suggest that the observed reduction in diversity of sequences generated after reintegration results from the exclusion of non-functional regions, effectively refining the functional sequence space in the $P^2$ model compared to the standard DCA model $P^1$.

\subsection{Experimental Validation on Group I Intron Ribozymes}

Group I intron ribozymes are catalytic RNA molecules that possess the ability to self-splice, meaning they can excise themselves from precursor RNA transcripts without the assistance of proteins or additional enzymes. Our work leverages on that of Lambert et al. \cite{Lambert2024-rd}, who started with a reference wildtype, the \emph{Azoarcus} group I intron ribozyme, and designed artificial mutations of this reference sequence using MSA-based generative models (DCA \cite{10.1093/nar/eaDCA} and VAE \cite{vae1}), structure based methods (Turner Model \cite{turner}) and combinations of the two. 
Through high-throughput assays, they tested these mutations for self-splicing-like activity and evaluated the percentage of active designs from each model at varying distances from the wild type.
Here, we present experimental evidence supporting the presented reintegration method applied in this context. \\
We start with their DCA model, utilizing it as our non-reintegrated $P^1$ model. This model was trained using as training set $\mathcal{D}_N$ an MSA of 817 group I introns aligned against the \emph{Azoarcus} group I intron ribozyme ($L=197$).
For the training of our reintegrated models, we employ as $\mathcal{D}_T$ a subset of 14099 from their experimentally labeled dataset, which consists of 24071 mutations of the reference (see \emph{Materials and Methods}).\\
An important detail is that the $\mathcal{D}_T$ in this case is significantly different from all the other reintegration instances in this study; all the tested designs are mutations of a single reference sequence, so their distribution in the sequence space is localized around one single point. Additionally, as expected, designs with fewer mutated residues generally exhibited higher activity than those with a greater number of mutations. Thus, a significant amount of information about sequence functionality contained in $\mathcal{D}_T$ is related to the trivial distance from the reference. Unlike all the other instances presented, where $\mathcal{D}_T$ designs are exclusively derived from the non-reintegrated model $P^1$, in this case, they originate from all the different models tested in \cite{Lambert2024-rd}. Details about $\mathcal{D}_T$ are provided in the \emph{Supplementary Section~S4} and \emph{Supplementary Figure~S3}. \\

We implemented two reintegration strategies:
\begin{enumerate}
    \item Standard Reintegration Procedure (REINT): All active $\mathcal{D}_T$ sequences ($\mathcal{D}_T^{+}$) were reintegrated with a weight $w(\underline{b}) = 1/|\mathcal{D}_T^+|$, and all non-active $\mathcal{D}_T$ sequences ($\mathcal{D}_T^{-}$) were reintegrated with $w(\underline{b}) = -1/|\mathcal{D}_T^-|$.
    \item Bin Sum Zero Reintegration (REINT BS0): $\mathcal{D}_T$ sequences were grouped into bins based on their mutational distance from the reference sequence, with each bin covering four mutational steps. For sequences in bin $i$, the weight $w(\underline{b})$ was set to $1/|b_i^+|$ for active sequences, where $|b_i^+|$ is the number of active sequences in the bin, and $w(\underline{b}) = -1/|b_i^-|$ for non-active sequences, where $|b_i^-|$ is the number of non-active sequences in the bin. This ensured that the sum of $w(\underline{b})$ in each bin was zero, mitigating bias towards the \emph{Azoarcus} reference sequence by balancing positive and negative signals at each distance.
\end{enumerate}
For further details and the choice of the parameter $\lambda$, refer to the \textit{Supplementary Section S4}. \\
The two models were trained using the same GPU DCA implementation used for the CM case.\\
From the two reintegrated models, we generated designs at various bins of mutational distance from the reference wildtype (Table 6). These designs were then experimentally tested for self-splicing activity using the same experimental assay that was used for the designs from the non-reintegrated $P^1$ model \cite{Lambert2024-rd}. Details
about the experimental procedures and its comparability across different experimental istances are provided in the \textit{Supplementary Section~S4} and \textit{Supplementary Figure~S4, S5} and \textit{Supplementary Table~S5}. \\

\begin{table}[h]
\centering

\begin{tabular*}{\columnwidth}{@{}l@{\hskip 11pt}c@{\hskip 11pt}c@{\hskip 11pt}c@{\hskip 11pt}c@{\hskip 11pt}c@{\hskip 11pt}c@{\hskip 11pt}c@{}}
\toprule
Model & 30 & 45 & 55 & 60 & 65 & 70 & 75 \\
\colrule
REINT    & -  & 80 & 180 & 180 & 180 & 180 & -  \\
REINT BS0 & 100 & 150 & - & 275 & - & - & 275 \\
\botrule
\end{tabular*}
\caption{\textit{Number of tested designs of the two reintegrated models at various mutational distances. For the DCA $P^1$ model, this number is 150 for each bin \cite{Lambert2024-rd}.}}
\label{table:dist}
\end{table}

The results of the experimental assays are displayed in Figure \ref{resultsGroup1} and detailed in the \textit{Supplementary Table~S6}. \\

\begin{figure}[h]
\begin{center}
\includegraphics[width=1\columnwidth]{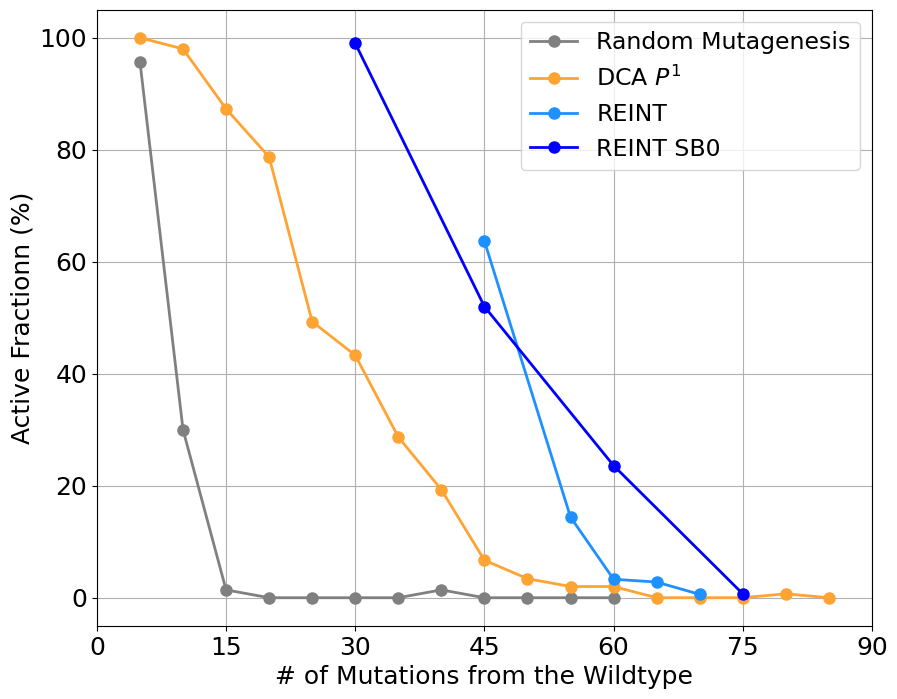}
\end{center}
\caption{\textit{Active fraction of the generated designs as a function of the mutational distance from the reference sequence for the $P^1$ DCA model and the two reintegrated models. The number of tested designs at each bin of distance can be found in Table \ref{table:dist}.}}
\label{resultsGroup1}
\end{figure}

Both reintegrated models outperformed the non-reintegrated one. The REINT BS0 model maintained an active fraction of 23.6\% at a mutational distance of 60 residues, compared to the non-reintegrated $P^1$ model, which had a similar active fraction of 19.3\% at 40 mutations but dropped to just 2.0\% at 60 mutations. The REINT model successfully generated functional sequences at 65 mutations, further away than all the $P^1(\underline{a})$ tested in \cite{Lambert2024-rd}, where the furthest active seqwuence was found at 60 mutations. \\
This increase in performance came at the expense of generated sequence diversity, which was significantly more impacted than in the previous reintegration examples. This is most likely due to the highly localized nature of the reintegrated dataset $\mathcal{D}_T$. The intra-sample and $\mathcal{D}_T^+$ distance distributions for mutational distances of 45, 60 and 65 are shown in Table \ref{table:Intra}, with the complete analysis provided in the \textit{Supplementary Table~S6} and \textit{Supplementary Figure~S6, S7}.

\begin{table}[h]
\centering

\begin{tabular*}{\columnwidth}{@{}l@{\hskip 22pt}c@{\hskip 22pt}c@{\hskip 22pt}c@{}}
\toprule
Model (Distance) & Active \% & \( D_{P^2-P^2} \) & \( D_{P^2-\mathcal{D}_T^+} \) \\
\colrule
DCA $P^1$ (45)   &  6.7 & 51.3 & 31.8 \\
REINT (45)       & 63.7 & 8.7 & 6.5 \\
REINT BS0 (45)   & 52.0& 10.0 & 7.6 \\
\botrule
DCA $P^1$ (60)   &  2.0 & 62.8 & 43.1 \\
REINT (60)       & 3.3 & 12.5 & 18.6 \\
REINT BS0 (60)   & 23.6 & 15.1 & 13.8 \\
\botrule
DCA $P^1$ (65)   &  0.0 & 70.4 & 50.3 \\
REINT (65)       &  2.8 & 18.1 & 28.0 \\
\botrule
\end{tabular*}
\caption{\textit{Active Fraction, Average Intra-Dataset Distance (\( D_{P^2-P^2} \)), and Average Minimum Distance from the Positively Reintegrated Dataset (\( D_{P^2-\mathcal{D}_T^+} \)) for mutational distances 60, 70, and 75 from the reference sequence.}}
\label{table:Intra}
\end{table}

\section{DISCUSSION}

Advances in high-throughput experimentation and machine learning are rapidly reshaping our ability to design functional biomolecular sequences. In this work, we have demonstrated that reintegrating experimental feedback into generative models markedly improves the reliability of predicted sequences. Our results show that even when the underlying mathematical model remains unchanged, the incorporation of well‐characterized experimental data – including both functional and non‐functional sequences – serves as a powerful corrective mechanism. Experimental tests therefore do not only validate predictions of models entirely trained on natural sequence data, but also guides the refinement of the modeled sequence space by penalizing non-functional regions and reinforcing areas associated with activity.

A key insight from our study is that enhanced performance can be achieved by training on more informative data rather than solely by increasing model complexity. The same DCA framework, when trained with a balanced integration of natural sequence data and experimental outcomes, produces a model that yields a significantly higher fraction of functional sequences. This observation underscores that the limitations of conventional generative models are not necessarily due to the inadequacy of their architectures, but rather stem from the sparsity and incomplete sampling inherent in natural sequence databases. By reintegrating experimental feedback, our method compensates for this deficiency, thereby sharpening the model’s discriminative power.

At the same time, our approach has inherent limitations. Because the reintegration procedure relies on experimental data that are necessarily sampled from regions already explored by nature, the model is not readily extended to completely novel areas of sequence space. In other words, while our method is highly effective at refining and navigating the known functional landscape, it remains dependent on the availability and quality of experimental measurements. This dependency emphasizes that the generation of truly novel sequences will continue to require a comprehensive experimental framework to guide and validate model predictions.

Looking ahead, the iterative interplay between experimental feedback and model refinement presents a promising route toward more accurate and targeted design strategies. Future studies could explore how successive rounds of data integration may gradually expand the functional sequence space, while also addressing potential trade-offs between accuracy and diversity. Ultimately, our results suggest that a synergistic integration of experimental and computational approaches can overcome the false-positive limitations of current generative models and pave the way for more reliable biomolecular design.

\section{MATERIALS AND METHODS}

In this section, we detail the methodological framework of our proposed reintegration method applied to the DCA generative model, as well as the data and methods employed for training and evaluating our models in a number of diverse RNA and protein families.

\subsection{Reintegration method for DCA models}
\label{sec:reintegration_dca}

The presented reintegration procedure is particularly well-suited for application to DCA models, as it does not require modifying their standard inference methods (differently from previous reintegration attempts \cite{Pierra}). In its conventional implementation, DCA assumes that the natural data distribution is described by a probability \(P^1(\underline{a})\) in the form of a Potts model,
\begin{equation}
P^1(\underline{a}) = \frac{1}{Z^1} \exp\Bigg\{ \sum_{i} h_i^1(a_i) \;+\; \sum_{i<j} J_{ij}^1(a_i,a_j) \Bigg\},
\label{eq:7}
\end{equation}
with \(\underline{a} = (a_1, \dots, a_L)\) denoting, according to the problem under study, an aligned nucleotide or amino-acid sequence of length \(L\). The optimal parameters \(\{h^1, J^1\}\) for \(P^1(\underline{a})\) are inferred via Maximum Likelihood Estimation (MLE) \cite{cuturello2020assessing},
\begin{equation}
\{h^1, J^1\} \;=\; \arg\max_{h,\,J} \; \mathcal{L}(h, J \,\vert\, \mathcal{D}_N)\,,
\label{eq:8}
\end{equation}
where \(\mathcal{D}_N\) is an MSA of all sequenced homologs of the family under consideration.  

We omit here the standard sequence reweighting procedure \cite{cocco2018inverse} commonly used in DCA, to simplify notation, but it can be included straightforwardly. Our publicly available implementation does contain it.
The reintegration data consist of a second alignment, \(\mathcal{D}_T\), of experimentally tested sequences, together with experimental outcomes encoded by the adjustment weight \(w(\underline{b})\). The reintegrated model \(P^2(\underline{a})\) remains in the Potts-model form,
\begin{equation}
P^2(\underline{a}) \;=\; \frac{1}{Z^2}\,\exp\Bigg\{\sum_{i}h_i^2(a_i)\;+\;\sum_{i<j}J_{ij}^2(a_i,a_j)\Bigg\},
\label{eq:9}
\end{equation}
but with updated parameters \(\{h^2, J^2\}\) chosen to maximize the new objective function
\begin{equation}
\{h^2, J^2\} \;=\; \arg\max_{h,\,J}\;\mathcal{Q}(h, J \,\vert\, \mathcal{D}_N, \mathcal{D}_T)\,.
\label{eq:10}
\end{equation}

In standard DCA inference, the MLE conditions require the one-point marginals \(P^1_i(a)\) and two-point marginals \(P^1_{ij}(a,b)\) of \(P^1(\underline{a})\) to match the empirical frequencies \(f_i(a)\) and \(f_{ij}(a,b)\) observed in \(\mathcal{D}_N\):
\begin{eqnarray}
\label{eq:marginal1}
P^1_i(a) &=&\frac{1}{|\mathcal{D}_N|} \sum_{\underline{a} \in \mathcal{D}_N} \delta_{a_i,a} =f_i(a) \\
P^1_{ij}(a,b) &=&\frac{1}{|\mathcal{D}_N|} \sum_{\underline{a},\underline{b} \in \mathcal{D}_N} \delta_{a_i,a} \cdot \delta_{a_j,b} =f_{ij}(a,b)
\nonumber
\end{eqnarray} 
where \(\delta_{a,b}\) is the Kronecker delta.

By contrast, maximizing \(\mathcal{Q}\) imposes choosing \(\{h^2, J^2\}\) so that the one-point and two-point marginals of \(P^2(\underline{a})\) match the \(\mathcal{D}_T\)-corrected frequencies \(\tilde{f}\) (see \emph{Supplementary Section S1}):
\begin{eqnarray}
\label{eq:corrected_freq1}
\tilde{f}_i(a) &=& \frac 1z \left[ f_i(a) + \dfrac{\lambda}{|\mathcal{D}_T|} \sum_{\underline{b} \in \mathcal{D}_T} w(\underline{b}) \cdot \, \delta_{b_i,a} \right]\\
\tilde{f}_{ij}(a,b) &=& \frac 1z \left[ f_{ij}(a,b) + \dfrac{\lambda}{|\mathcal{D}_T|} \sum_{\underline{b} \in \mathcal{D}_T} w(\underline{b}) \cdot \, \delta_{b_i,a} \cdot \delta_{b_j,b} \right]
\nonumber
\end{eqnarray} 
where 
\begin{equation}
z \;=\; 1 \;+\;\dfrac{\lambda}{|\mathcal{D}_T|}\,\sum_{\underline{b}\in\mathcal{D}_T}\,w(\underline{b}).
\end{equation}
Hence, the maximization condition becomes
\begin{eqnarray}
    \label{eq:Obj1}
P^2_i(a) &=& \tilde{f}_i(a) \nonumber\\
P^2_{ij}(a,b)  &=& \tilde{f}_{ij}(a,b)
\end{eqnarray} 
for all $i,j$ and all $a,b$. Thus, any standard DCA inference method may be used simply by replacing the original empirical frequencies \(f\) with the corrected frequencies \(\tilde{f}\) from Eq.~(\ref{eq:corrected_freq1}). Consequently, implementing reintegration in a DCA pipeline requires only altering the frequency targets used during parameter inference.

It is important to note, however, that negative values of \(w(\underline{b})\) (i.e.\ experimentally tested non-functional sequences) can break the convexity of the problem, removing guaranteed uniqueness or even existence of a solution.  Additionally, the corrected frequencies \(\tilde{f}\) may sometimes lie outside the interval \([0,1]\).  As a result, convergence to a consistent DCA model is not strictly guaranteed, especially for strong reintegration strengths~\(\lambda\). Nonetheless, in all of our applications, we found that moderate settings of~\(\lambda\) do converge reliably.

\subsection{Data, RNA fitness proxy \& Group I intron experimental activity}\label{data}

\subsubsection{RFAM sequence data} -- The RNA datasets used in this paper are MSA of three RNA families. These families are the tRNA family RF00005 (number of sequences $|\mathcal{D}_N|=30000$, number of residues $L=71$), the SAM riboswitch family RF0162 ($|\mathcal{D}_N|=6113$, $L=108$), and the Glycine riboswitch family RF0504 ($|\mathcal{D}_N|=4600$, $L=94$). For all three RNA families, the corresponding consensus secondary structure is available from the Rfam database \cite{Rfam}. The Datasets are taken from \cite{10.1093/nar/eaDCA}.

\subsubsection{RNA fitness proxy} -- To study the reintegration procedure, we need a fast and reliable way to label our datasets. Performing experiments on RNA sequence functionality is both expensive and time-consuming, so we initially used computational tools as fitness proxies to label our datasets. Since we have access to the RNA families' consensus secondary structures, we employed the  \texttt{RNAeval} function provided by the Vienna Package \cite{ViennaRNA}. \texttt{RNAeval} calculates the thermodynamic free energy \( F \) (using the Turner 2004 energy model \cite{turner}) of an RNA sequence folded onto a given secondary structure.

Our underlying assumption is that well-generated sequences will, on average, exhibit low free energy \( F \) when folded onto the family's consensus structure, indicating structural stability. In contrast, poorly generated sequences may lack structural stability or fold into alternative structures, and are expected to have a higher free energy. Therefore, we will use \( -F \) as a fitness proxy. This approach leverages the established structure/function relationship in RNA \cite{structure_function, structure_function2}, and enables us to computationally label sequences for a given target secondary structure.

\subsubsection{Group I Intron Ribozymes Data}

-- The Group I intron datasets utilized in this paper are taken from the study by Lambert et al. \cite{Lambert2024-rd}. These datasets consist of multiple sequence alignments (MSAs) of Group I intron ribozymes. The natural data \( D_N \) consists of an MSA of \( N = 817 \) natural Group I introns aligned against the reference sequence, the Azoarcus Group I intron (numbeer of residues \( L = 197 \)). The reintegration dataset \( D_T \) consists of a subset of 14099 out of the original 24071 experimentally labeled sequences. Specifically, \( D_T \) includes the sequences with mutational distance from 3 to 60, excluding those near the activity threshold. Sequences with mutational distances exceeding 60 were not reintegrated due to the unreliable detection of activity signals beyond this range (see \emph{Supplementary Section S4}).

\subsubsection{Group I Experimental Activity}

-- To experimentally validate the reintegration methods, we conducted experiments on artificially designed Group I intron ribozymes. The experimental activity was determined using the same methodology as the tested ribozymes in Lambert et al.~\cite{Lambert2024-rd}. The experimental assay consists of a high-throughput screening of self-splicing catalytic activity, details are provided in the \emph{Supplementary Section S4}.

\subsubsection{Protein data} -- The protein datasets used in this paper are taken from the study by Russ et al. \cite{russetal}. These datasets consist of MSAs of the chorismate mutase (CM) enzyme. $\mathcal{D}_N$ consists of the MSA of the natural CM homologs (number of sequences $N=1130$, number of residues $L=96$) and the reintegration dataset $\mathcal{D}_T$ is the alignment ($N=1003$, $L=96$) of DCA-generated artificial CM variants. In \cite{russetal}, both datasets are labeled based on experimental testing: all protein sequences were expressed in genetically engineered \textit{E. coli} strains, each modified to produce one of the CMs from $\mathcal{D}_N$ or $\mathcal{D}_T$ instead of their natural wildtype variants. These \textit{E. coli} strains were then tested for growth under selective conditions; sequences enabling \textit{E. coli} growth were labeled as functional, whereas those that did not were labeled as non-functional. It is noteworthy that many natural sequences in $\mathcal{D}_N$ were non-functional in \textit{E. coli} under the experiment's conditions:  the natural CM homologs have undergone evolutionary adaptations specific to their native hosts, and may fail to provide growth in the \textit{E. coli} environment.
\\

\section*{Data and Code Availability}

All data and code used in this study are publicly available at Zenodo:  
\href{https://doi.org/10.5281/zenodo.15115193}{10.5281/zenodo.15115193}.

\section*{ACKNOWLEDGEMENTS}

We acknowledge helpful discussions with Camille Lambert, Roberto Netti, Vaitea Opuu, Andrea Pagnani, Lorenzo Rosset and Francesco Zamponi during the preparation of this work, and we thank particularly Lorenzo Rosset for providing his GPU-based adabmDCA implementation prior to publication, and to Vaitea Opuu for help with the initial analysis of the experimental data using the scripts of \cite{Lambert2024-rd}.

\subsubsection{Conflict of interest statement.} None declared.

%% file: parts/Supplement.tex
\clearpage
\onecolumngrid

\renewcommand{\thesection}{S\arabic{section}}
\renewcommand{\thesubsection}{\thesection.\arabic{subsection}}
\renewcommand{\thesubsubsection}{\thesubsection.\arabic{subsubsection}} 

\renewcommand{\thetable}{S\arabic{table}}
\renewcommand{\thefigure}{S\arabic{figure}}
\renewcommand{\theequation}{S\arabic{equation}}

\setcounter{section}{0}
\setcounter{table}{0}
\setcounter{figure}{0}
\setcounter{equation}{0}

\begin{center}
    \Huge \textbf{Supplementary Information}
\end{center}

\section{DCA Analytical Computations}
$ P^2(\underline{a})$ is a Boltzmann distribution over the sequence space,
\begin{equation} \label{eq:Boltz}
    P^2(\underline{a}) =\frac{1}{Z^2} \exp\big\{-H^2(\vec{a})\big\}\ ,
\end{equation}
defined via the Potts Hamiltonian ${H^2(\vec{a})}$
\begin{equation} \label{eq:Ham}
    H^2(\underline{a}) =\sum_{i}h^2_i(a_i) + \sum_{i<j}J^2_{ij}(a_i,a_j) \ .
\end{equation}
The optimal parameters $\{ h^2, J^2\} $ are obtained via
\begin{equation} \label{eq:MLEcond}
    \{ h^2, J^2\} = \underset{h, J}{\text{argmax}} \text{\space} \mathcal{Q}(h, J | \mathcal{D}_N, \mathcal{D}_T)
\end{equation}
as arguments of the maximum of the new objective function
\begin{equation} \label{eq:new_objective}
Q( h, J | \mathcal{D}_N, \mathcal{D}_T) =\frac{1}{|\mathcal{D}_N|} \sum_{\underline{a} \in  \mathcal{D}_N} \ln P^2(\underline{a} | h, J) + \frac{\lambda}{|\mathcal{D}_T|}\sum_{\underline{b}\in  \mathcal{D}_T} w(\underline{b}) \cdot \ln P^2(\underline{b} | h, J) 
\ .
\end{equation}
It is possible to exploit Eq.~\eqref{eq:Boltz} to write :
\begin{equation*}
    {\cal Q} = - \frac{1}{|\mathcal{D}_N|}   \sum_{\underline{a} \in {\cal D}_N} H^2(\underline{a}) - \log Z^2
    -\frac{\lambda}{|\mathcal{D}_T|} \sum_{\underline{b} \in {\cal D}_T} H^2(\underline{b})\cdot w(\vec{b}) 
    - \frac{\lambda}{|\mathcal{D}_T|} \log Z^2  \sum_{\underline{b} \in {\cal D}_T} w(\underline{b})  \ .
\end{equation*}
The aim is to analytically maximize $Q( h, J | \mathcal{D}_N, \mathcal{D}_T)$ to find $\{ h^2, J^2\} $. To this end, we need to compute partial derivatives of $Q$ with respect to fields $h^2_i(a)$ and couplings $J_{ij}^2(a,b)$. Proceeding term by term we find
\begin{eqnarray}\label{eq:delta}
    -\frac{1}{|\mathcal{D}_N|} \sum_{\underline{a} \in {\cal D}_N} \frac{\partial H^2(\underline{a})}{\partial h_i^2(a)} &=&-\frac{1}{|\mathcal{D}_N|} \sum_{\underline{a} \in {\cal D}_N} \frac{\partial \big( \sum_{j}h^2_j(a_j) \big)  }{\partial h_i^2(a)} \\
    &=& -\frac{1}{|\mathcal{D}_N|} \sum_{\underline{a} \in {\cal D}_N}  \delta_{a_i,a} \nonumber\\
    &=&f_i(a) \ . \nonumber
\end{eqnarray}
Similarly we obtain
\begin{equation*}
    -\frac{1}{|\mathcal{D}_N|} \sum_{\underline{a} \in {\cal D}_N} \frac{\partial H^2(\underline{a})}{\partial J_{ij}^2(a,b)} =  f_{ij}(a,b) \ .
\end{equation*}
The derivatives of the logarithm of partition function can be computed exploiting the same reasoning of Eq.~\eqref{eq:delta},
\begin{eqnarray}
    -\frac{\partial \log Z^2}{\partial h_i^2(a)} &=& \frac{1}{Z^2} \frac{\partial Z^2}{\partial h_i^2(a)} = \frac{1}{Z^2} \sum_{\underline{a}}  \frac{\partial }{\partial h_i^2(a)} \exp\big\{-H^2(\underline{a})\big\}\\ &=&
  \frac{1}{Z^2} \sum_{\underline{a}}  \delta_{a_i,a} \cdot \exp\big\{-H^2(\underline{a})\big\} \nonumber\\
    &=& \sum_{\underline{a}}  \delta_{a_i,a} \cdot P^2(\vec{a})=  P_i^2(a) \ . \nonumber
\end{eqnarray}
Similarly we obtain
\begin{equation*}
    -\frac{\partial \log Z^2}{\partial J_{ij}^2(a,b)} =  P_{ij}^2(a,b) \ .
\end{equation*}
Finally, using again Eq.~\eqref{eq:delta}, the derivatives of terms involving the adjustment function $w(\vec{b})$ can be computed,
\begin{eqnarray}
    -\sum_{\underline{b} \in {\cal D}_T} \frac{\partial H^2(\underline{b})\cdot w(\underline{b})}{\partial h_i^2(a)} &=&  \sum_{\underline{b} \in {\cal D}_T}  \delta_{b_i,a} \cdot w(\underline{b}) 
    \\
    -\sum_{\underline{b} \in {\cal D}_T} \frac{\partial H^2(\underline{b})\cdot w(\underline{b})}{\partial J_{ij}^2(a,b)}
    &=&  \sum_{\underline{b} \in {\cal D}_T}  \delta_{b_i,a} \cdot  \delta_{b_j,b} \cdot w(\underline{b})\nonumber
\end{eqnarray}
Rearranging terms, the following equations for the first and second moment of $P^{2}(\vec{a})$ are found: 
\begin{eqnarray}\label{eq:corrected_freq1}
P_i^2(a) &=& \frac 1z \left[ f_i(a) + \dfrac{\lambda}{|\mathcal{D}_T|} \sum_{\underline{b} \in \mathcal{D}_T} w(\underline{b}) \cdot \, \delta_{b_i,a} \right] = \tilde{f}_i(a)\\
P_{ij}^2(a,b) &=& \frac 1z \left[ f_{ij}(a,b) + \dfrac{\lambda}{|\mathcal{D}_T|} \sum_{\underline{b} \in \mathcal{D}_T} w(\underline{b}) \cdot \, \delta_{b_i,a} \cdot \delta_{b_j,b} \right] = \tilde{f}_{ij}(a,b)
\nonumber
\end{eqnarray} 
with normalization
\begin{equation}
z = {1 + \dfrac{\lambda}{|\mathcal{D}_T|} \sum_{\vec{b} \in \mathcal{D}_T} w(\underline{b})}  \ .
\end{equation}
The model training can be performed using all the standard DCA training techniques, using the adjusted frequencies $\tilde f_i$ and $\tilde f_{ij}$ in Eq.~\ref{eq:corrected_freq1} as targets for the model's marginals instead of the empirical MSA frequencies $f_i$ and $f_{ij}$.

\section{RNA RFAM families}
\subsection{Edge Activation DCA Training }

For each RFAM family, we used the Edge Activation DCA (eaDCA) algorithm from \cite{10.1093/nar/eaDCA} to train a DCA model \( P^1(\underline{a}) \) on the MSA of the RNA family. We used 8000 chains for training and a pseudocount of 0.05. The empirical frequencies $f_i(a)$ and $f_{ij}(a,b)$ are computed along with the correlation matrix $C_{ij}^{emp}(a,b)= f_{ij}(a,b)-f_{i}(a)f_{j}(b)$. Training stops when the Pearson correlation between $C_{ij}^{emp}(a,b)$ and $C_{ij}^{train}(a,b)$ reaches 0.95. \\
Once the model is learned, we sample artificial sequences with Gibbs Sampling, forcing the obtained sequences to have no gaps. This is done to ensure that the \texttt{RNAeval} proxy fitness, $-F$ is comparable across different sequences. A dataset of 2000 artificial sequences was sampled from each model. These sequences, along with a \texttt{RNAeval} proxy fitness $-F$ for each of them, serve as Reintegration Dataset $\mathcal{D}_T$ for our reintegration procedure. 

Subsequently, we computed the effective frequencies using Equation~\ref{eq:corrected_freq1} to train the reintegrated model \( P^2(\underline{a}) \). Depending on the value of $\lambda$, the convergence of the reintegrated model’s training is not always guaranteed. To facilitate convergence of the eaDCA procedure, we set any negative effective frequencies to zero and enforced a constraint to prevent the same edge from being activated more than five times. All other training settings were kept identical to those used for training the non-reintegrated \( P^1(\underline{a}) \) model.

 \subsection{Results for different values of $\lambda$ }

We tested the reintegration procedure at different values of \(\lambda\), starting with \(\lambda = 0\) and increasing \(\lambda\) by 0.1 at each step. The maximum \(\lambda\) tested corresponds to the largest value below 2 for which our new model converges within \(10^4\) steps using the eaDCA algorithm. For RF00504, the resulting interval of tested \(\lambda\) values is \([0.1,1.4]\), for RF00162 \([0.1,1]\), and for RF00005 \([0.1:2]\). The complete results of these analyses are presented for all three families in Tables ~\ref{table1},~\ref{table2}, and~\ref{table3}.

Furthermore, we provide, for each family, the histograms of \texttt{RNAeval} \cite{ViennaRNA} proxy fitness $-F$ for \( \lambda= 0.1,\, 0.5\) and $\lambda = \lambda _{max}$. These can be found in Figure 1.
\begin{figure}[H]
\begin{center}
\includegraphics[width=\columnwidth]{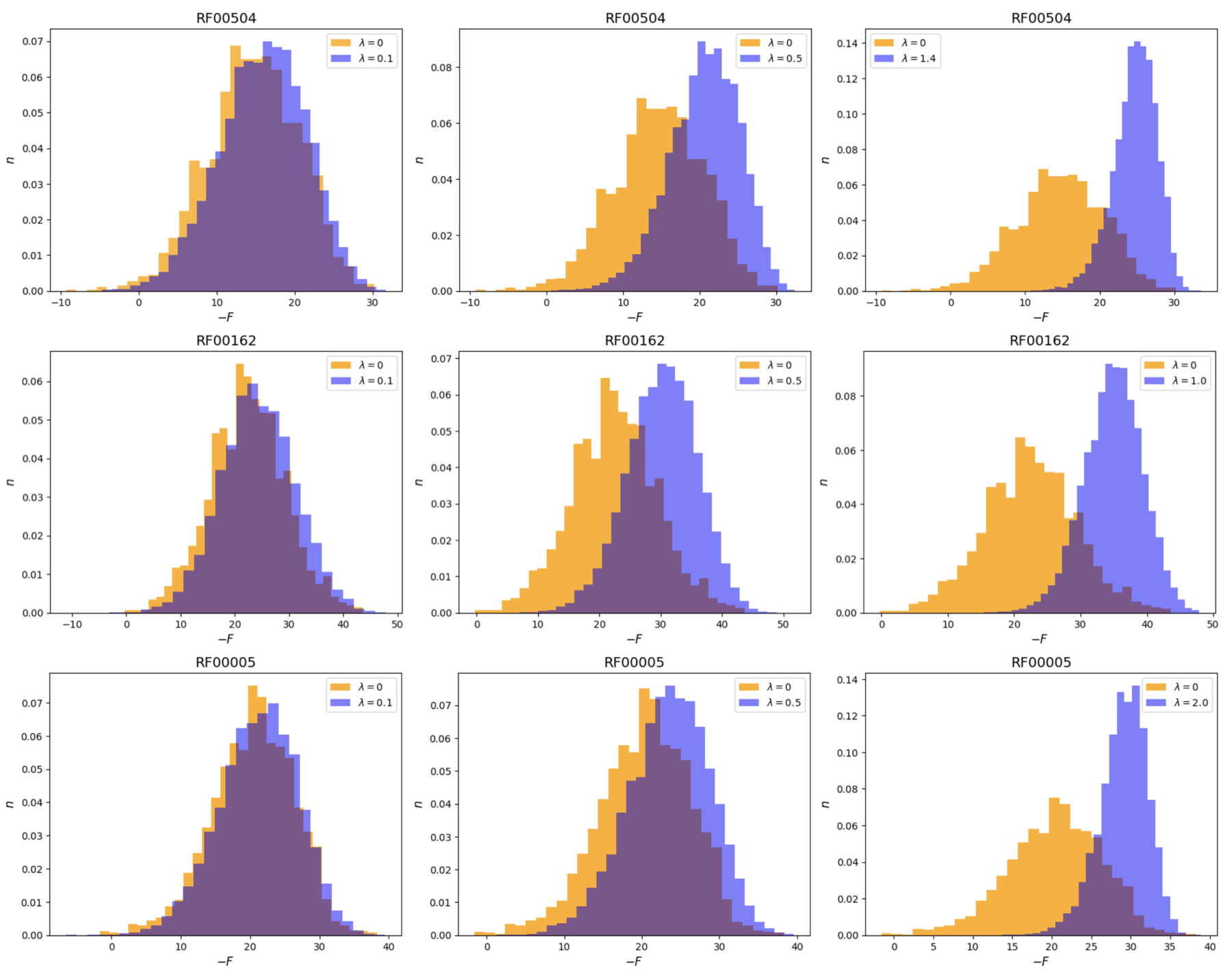}
\end{center}
\caption{\\
\textit{Distribution of the \texttt{RNAeval} proxy fitness for the RF00504 DCA model \( P^1(\underline{a}) \) (orange, \( \lambda = 0 \)) and the reintegrated \( P^2(\underline{a}) \) model (blue, \( \lambda = 0.1 \), \( \lambda = 0.5 \), \( \lambda \)= 1.4)\\
Distribution of the \texttt{RNAeval} proxy fitness for the RF00162 DCA model \( P^1(\underline{a}) \) (orange, \( \lambda = 0 \)) and the reintegrated \( P^2(\underline{a}) \) model (blue, \( \lambda = 0.1 \), \( \lambda = 0.5 \), \( \lambda \)= 1.0)\\ 
Distribution of the \texttt{RNAeval} proxy fitness for the RF00005 DCA model \( P^1(\underline{a}) \) (orange, \( \lambda = 0 \)) and the reintegrated \( P^2(\underline{a}) \) model (blue, \( \lambda = 0.1 \), \( \lambda = 0.1 \), \( \lambda \)= 2.0)}}
\label{histograms}
\end{figure}
\begin{table}
\begin{tabular*}{\columnwidth}{@{\extracolsep{\fill}}lllllll@{}}
\toprule
$\lambda$ value & \hskip 5pt TP & \hskip 10pt \(-\tilde{F}\) \hskip 10pt & \hskip 6pt $S$ & \hskip 2pt \( D_{P^2-P^2} \)  & \hskip 2pt \( D_{P^2-\mathcal{D}_T^+} \)  & \hskip 2pt \( D_{P^2-\mathcal{D}_T^-} \) \hskip 10pt  \\
\midrule
\hskip 12pt$0.0$ & 51.6\%  &\hskip 12pt 14.7  & 61.9 & \hskip 12pt 43.3  & \hskip 21pt -- & \hskip 21pt -- \\
\hskip 12pt$0.1$ & 60.3\%  & $15.9 \pm 0.1$ & 63.8 & $42.8 \pm 0.1$ & $26.0 \pm 0.1$ & $27.8 \pm 0.1$ \\
\hskip 12pt$0.2$ & 69.6\%  & $17.2 \pm 0.1$ & 63.3 & $41.2 \pm 0.1$ & $25.0 \pm 0.1$ & $27.5 \pm 0.1$     \\
\hskip 12pt$0.3$ & 80.6\%  & $18.9 \pm 0.1$ & 62.9 & $39.4 \pm 0.1$ & $24.2 \pm 0.1$ & $27.3 \pm 0.1$  \\
\hskip 12pt$0.4$ & 84.5\%  & $19.6 \pm 0.1$ & 62.6 & $38.5 \pm 0.1$ & $23.8 \pm 0.1$ & $27.2 \pm 0.1$  \\
\hskip 12pt$0.5$ & 89.6\%  & $20.6 \pm 0.1$ & 61.2 & $37.3 \pm 0.1$ & $23.3 \pm 0.1$ & $27.1 \pm 0.1$   \\
\hskip 12pt$0.6$ & 93.3\%  & $21.7 \pm 0.1$ & 60.6 & $35.6 \pm 0.1$ & $22.7 \pm 0.1$ & $26.8 \pm 0.1$  \\
\hskip 12pt$0.7$ & 96.1\%  & $22.5 \pm 0.1$ & 60.1 & $34.7 \pm 0.2$ & $22.2 \pm 0.1$ & $26.6 \pm 0.1$ \\
\hskip 12pt$0.8$ & 95.8\%  & $22.4 \pm 0.1$ & 58.7 & $34.1 \pm 0.1$ & $21.9 \pm 0.1$ & $26.4 \pm 0.1$    \\
\hskip 12pt$0.9$ & 97.6\%  & $23.1 \pm 0.1$ & 57.7 & $33.7 \pm 0.1$ & $21.7 \pm 0.1$ & $26.4 \pm 0.1$   \\
\hskip 12pt$1.0$ & 98.6\%  & $23.7 \pm 0.1$ & 55.9 & $33.2 \pm 0.1$ & $21.4 \pm 0.1$ & $26.2 \pm 0.1$  \\
\hskip 12pt$1.1$ & 98.7\%  & $23.8 \pm 0.1$ & 53.2 & $32.7 \pm 0.1$ & $21.2 \pm 0.1$ & $26.1 \pm 0.1$    \\
\hskip 12pt$1.2$ & 99.3\%  & $24.3 \pm 0.1$ & 51.6 & $32.5 \pm 0.1$ & $21.0 \pm 0.1$ & $26.1 \pm 0.1$  \\
\hskip 12pt$1.3$ & 99.4\%  & $24.5 \pm 0.1$ & 48.3 & $32.0 \pm 0.1$ & $20.6 \pm 0.1$ & $25.9 \pm 0.1$ \\
\hskip 12pt$1.4$ & 99.7\%  & $24.7 \pm 0.1$ & 43.3 & $31.6 \pm 0.1$ & $20.3 \pm 0.1$ & $25.7 \pm 0.1$  \\
\bottomrule
\end{tabular*}
\caption{RF504}
\label{table1}
\end{table}
\begin{table}
\begin{tabular*}{\columnwidth}{@{\extracolsep{\fill}}lllllll@{}}
\toprule
$\lambda$ value & \hskip 5pt TP & \hskip 10pt \(-\tilde{F}\) \hskip 10pt & \hskip 6pt $S$ & \hskip 2pt \( D_{P^2-P^2} \)  & \hskip 2pt \( D_{P^2-\mathcal{D}_T^+} \)  & \hskip 2pt \( D_{P^2-\mathcal{D}_T^-} \) \hskip 10pt  \\
\midrule
\hskip 12pt$0.0$ & 49.9\%  & \hskip 12pt 22.4  & 63.5 & \hskip 12pt 46.0  & \hskip 21pt -- & \hskip 21pt-- \\
\hskip 12pt$0.1$ & 60.4\%  & $24.3 \pm 0.2$ & 65.3  & $45.8 \pm 0.1$ & $28.1 \pm 0.1$ & $29.6 \pm 0.1$ \\
\hskip 12pt$0.2$ & 69.5\%  & $25.9 \pm 0.1$ & 65.6  & $44.9 \pm 0.1$ & $27.7 \pm 0.1$ & $29.8 \pm 0.1$ \\
\hskip 12pt$0.3$ & 80.0\%  & $27.7 \pm 0.1$ & 64.7  & $43.8 \pm 0.2$ & $27.2 \pm 0.1$ & $30.0 \pm 0.1$ \\
\hskip 12pt$0.4$ & 85.2\%  & $28.9 \pm 0.1$ & 63.9  & $42.9 \pm 0.1$ & $27.0 \pm 0.1$ & $30.3 \pm 0.1$ \\
\hskip 12pt$0.5$ & 91.3\%  & $30.2 \pm 0.2$ & 63.2  & $42.2 \pm 0.1$ & $26.7 \pm 0.1$ & $30.5 \pm 0.1$ \\
\hskip 12pt$0.6$ & 94.7\%  & $31.3 \pm 0.1$ & 62.3  & $41.6 \pm 0.1$ & $26.4 \pm 0.1$ & $30.6 \pm 0.1$ \\
\hskip 12pt$0.7$ & 96.6\%  & $32.2 \pm 0.1$ & 61.4  & $40.9 \pm 0.1$ & $26.2 \pm 0.1$ & $30.9 \pm 0.1$ \\
\hskip 12pt$0.8$ & 98.0\%  & $33.0 \pm 0.1$ & 59.4  & $40.3\pm 0.1$ & $25.9  \pm 0.1$ & $31.0 \pm 0.1$ \\
\hskip 12pt$0.9$ & 98.9\%  & $33.8 \pm 0.1$ & 57.1  & $39.5 \pm 0.1$ & $25.6 \pm 0.1$ & $31.0 \pm 0.1$ \\
\hskip 12pt$1.0$ & 99.4\%  & $34.9 \pm 0.1$ & 53.3  & $39.7 \pm 0.1$ & $25.3 \pm 0.1$ & $31.8 \pm 0.1$ \\
\bottomrule
\end{tabular*}
\caption{RF162}
\label{table2}
\end{table}
\begin{table}
\begin{tabular*}{\columnwidth}{@{\extracolsep{\fill}}lllllll@{}}
\toprule
$\lambda$ value & \hskip 5pt TP & \hskip 10pt \(-\tilde{F}\) \hskip 10pt & \hskip 6pt $S$ & \hskip 2pt \( D_{P^2-P^2} \)  & \hskip 2pt \( D_{P^2-\mathcal{D}_T^+} \)  & \hskip 2pt \( D_{P^2-\mathcal{D}_T^-} \) \hskip 10pt  \\
\midrule
\hskip 12pt $0.0$ & 52.3\%  & \hskip 12pt 20.4  & 50.3 & \hskip 12pt36.3  & \hskip 21pt -- & \hskip 21pt -- \\
\hskip 12pt$0.1$ & 57.7\%  & $21.2 \pm 0.2$ & 51.2  & $36.1 \pm 0.1$ & $20.7 \pm 0.1$ & $22.2 \pm 0.1$ \\
\hskip 12pt$0.2$ & 61.6\%  & $21.7 \pm 0.1$ & 50.9  & $35.8 \pm 0.1$ & $20.5 \pm 0.1$ & $22.1 \pm 0.1$ \\
\hskip 12pt$0.3$ & 65.6\%  & $22.3 \pm 0.1$ & 50.2  & $35.7 \pm 0.1$ & $20.4 \pm 0.1$ & $22.1 \pm 0.1$ \\
\hskip 12pt$0.4$ & 69.2\%  & $22.9 \pm 0.2$ & 50.0  & $35.5 \pm 0.1$ & $20.2 \pm 0.1$ & $22.0 \pm 0.1$ \\
\hskip 12pt$0.5$ & 73.3\%  & $23.5 \pm 0.1$ & 49.5  & $35.3 \pm 0.1$ & $20.0 \pm 0.1$ & $22.0 \pm 0.1$ \\
\hskip 12pt$0.6$ & 76.4\%  & $23.9 \pm 0.2$ & 49.0  & $35.0 \pm 0.1$ & $19.8 \pm 0.1$ & $21.9 \pm 0.1$ \\
\hskip 12pt$0.7$ & 80.1\%  & $24.4 \pm 0.2$ & 48.9  & $34.8 \pm 0.1$ & $19.7 \pm 0.1$ & $21.9 \pm 0.1$ \\
\hskip 12pt$0.8$ & 83.3\%  & $24.9 \pm 0.1$ & 47.6  & $34.6 \pm 0.1$ & $19.5 \pm 0.1$ & $21.9 \pm 0.1$ \\
\hskip 12pt$0.9$ & 86.0\%  & $25.4 \pm 0.1$ & 47.6  & $34.3 \pm 0.1$ & $19.3 \pm 0.1$ & $21.9 \pm 0.1$ \\
\hskip 12pt$1.0$ & 88.6\%  & $25.7 \pm 0.1$ & 46.6  & $34.2 \pm 0.1$ & $19.3 \pm 0.1$ & $21.9 \pm 0.1$ \\
\hskip 12pt$1.1$ & 91.2\%  & $26.2 \pm 0.1$ & 46.1  & $34.1 \pm 0.1$ & $19.1 \pm 0.1$ & $21.9 \pm 0.1$ \\
\hskip 12pt$1.2$ & 93.9\%  & $26.8 \pm 0.1$ & 45.3  & $33.6 \pm 0.1$ & $18.9 \pm 0.1$ & $21.7 \pm 0.1$ \\
\hskip 12pt$1.3$ & 94.7\%  & $27.0 \pm 0.1$ & 44.4  & $33.5 \pm 0.1$ & $18.8 \pm 0.1$ & $21.7 \pm 0.1$ \\
\hskip 12pt$1.4$ & 96.0\%  & $27.4 \pm 0.1$ & 43.8  & $33.4 \pm 0.1$ & $18.7 \pm 0.1$ & $21.7 \pm 0.1$ \\
\hskip 12pt$1.5$ & 97.6\%  & $27.8 \pm 0.1$ & 42.8  & $33.3 \pm 0.1$ & $18.6 \pm 0.1$ & $21.8 \pm 0.1$ \\
\hskip 12pt$1.6$ & 98.1\%  & $28.2 \pm 0.1$ & 41.5  & $32.9 \pm 0.1$ & $18.4 \pm 0.1$ & $21.7 \pm 0.1$ \\
\hskip 12pt$1.7$ & 98.4\%  & $28.3 \pm 0.1$ & 39.8  & $32.8 \pm 0.1$ & $18.3 \pm 0.1$ & $21.7 \pm 0.1$ \\
\hskip 12pt$1.8$ & 98.8\%  & $28.5 \pm 0.1$ & 38.8  & $32.7 \pm 0.1$ & $18.2 \pm 0.1$ & $21.8 \pm 0.1$ \\
\hskip 12pt$1.9$ & 99.0\%  & $28.8 \pm 0.1$ & 36.5  & $32.3 \pm 0.1$ & $18.0 \pm 0.1$ & $21.7 \pm 0.1$ \\
\hskip 12pt$2.0$ & 99.1\%  & $29.1 \pm 0.1$ & 33.7  & $32.3 \pm 0.1$ & $18.0 \pm 0.1$ & $21.8 \pm 0.1$ \\
\bottomrule
\end{tabular*}
\caption{RF005}
\label{table3}
\end{table}

\newpage

\subsection{Fitness landscape prediction}
Previous studies \cite{contactsAndFitness,fitnessprediction2} suggest there should be an anti-correlation between the energy assigned by the model and the actual fitness of a sequence. In this section, we examine whether our experiment-informed reintegrated model improves its ability to predict sequence fitness.\\
Using each of the RNA families RF00504, RF00005, and RF00162, we generated datasets \(D_{\text{global}}\), consisting of 2000 independent sequences sampled from the non-reintegrated \(P^1(\underline{a})\) model with Gibbs Sampling.\\
Since \(P^1(\underline{a})\) was trained on the Natural MSA, these sequences display diversity similar to that of their respective natural RNA families and are widely distributed across the sequence space (Table~\ref{table1}, \ref{table2}, \ref{table3}). \\
For each sequence in \(D_{\text{global}}\), we calculated the energies of both the \(P^1(\underline{a})\) and \(P^2(\underline{a})\) models and measured their correlation with the proxy fitness \(-F\). Across all three RNA families, we observe a substantial increase in the correlation between model energy and proxy fitness after reintegration. This improvement depends on the reintegration strength parameter \(\lambda\), with higher \(\lambda\) values leading to stronger correlation enhancements. This can be seen in Figure \ref{Correlations Global}.

\begin{figure}[H]
\begin{center}
\includegraphics[width=0.8\columnwidth]{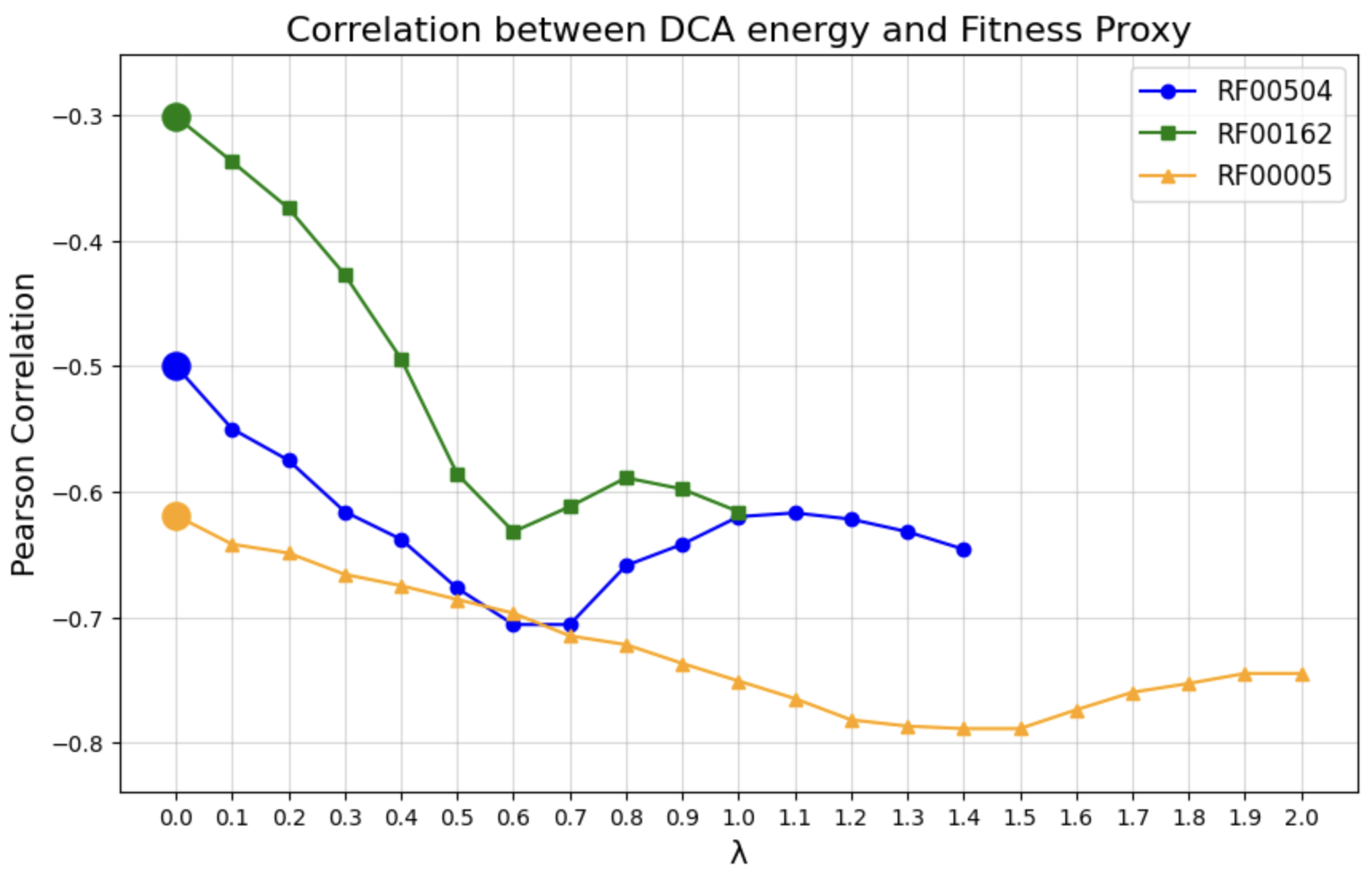}
\end{center}
\caption{\textit{Correlation between DCA energy and fitness proxy for three RNA families (RF00504, RF00162, RF00005) across the considered range of $\lambda$ values. At $\lambda=0$ (without reintegration), the starting points for each family are highlighted.}}
\label{Correlations Global}
\end{figure}
\newpage
\section{Chorismate Mutase}

\subsection{Classifier prediction for CM}

As described in \cite{russetal}, a logistic regression classifier is used to predict sequence functionality. The training set for the logistic regression consists of the experimentally labeled natural homologs (\(\mathcal{D}_N\)) published in \cite{russetal}. Each sequence is assigned a binary label \(x\), where \(x = 1\) denotes a sequence found to be functional under experimental assay conditions, and \(x = 0\) corresponds to an experimentally non-functional sequence. The predicted probability of a sequence $\underline a$ to be functional is given by 
\[
P(x = 1\, |\, \underline a) \sim \exp \left\{ g + \sum_{i=1}^{L} K_i(a_i) \right\},
\]
where \(g\) represents a bias term, and \(K_i(a_i)\) links the functionality \(x\) to the specific amino acid \(a_i\) at position \(i\).

The accuracy of this classifier, when tested on the \(\mathcal{D}_T\) protein dataset, is about 80\%.

\subsection{Training Procedure and Results for different values of $\lambda$ }

To apply the reintegration procedure on the Chorismate Mutase protein family, we used the Adaptive Boltzmann Machine DCA (adabmDCA) algorithm to train a DCA model \( P^2(\underline{a}) \) using the effective frequencies computed at different values of \( \lambda \). Depending on the value of $\lambda$, the convergence of the reintegrated model’s training is not always guaranteed. To facilitate convergence of the adabmDCA procedure, we set any negative effective frequencies to zero. All other training settings were kept identical to those used for training the non-reintegrated \( P^1(\underline{a}) \) model. The training was performed with 10,000 Monte Carlo chains, 10 sweeps per gradient update, and a learning rate of 0.05. Training stopped once the Pearson correlation between the empirical and training correlation matrices, \( C_{ij}^{emp}(a,b) \) and \( C_{ij}^{train}(a,b) \), reached 0.95.
The results of the reintegration procedure for \( \lambda = \{0.25, 0.5, 0.75, 1.0\} \) are shown in Table~\ref{table4}.  

\begin{table}[H]
\begin{tabular*}{\columnwidth}{@{\extracolsep{\fill}}lllll@{}}
\toprule
$\lambda$ value & \hskip -22pt Working (\%)  & \hskip -6pt \( D_{P^2-P^2} \)  & \hskip -6pt \( D_{P^2-\mathcal{D}_T^+} \)  & \hskip -6pt \( D_{P^2-\mathcal{D}_T^-} \) \hskip 10pt  \\
\midrule
$\lambda=0$   & \hskip -10pt 36.4 & \hskip 1pt 72.9 & \hskip 2pt -- & \hskip 2pt -- \hskip 10pt \\
$\lambda=0.25$  & \hskip -10pt 43.6  & \hskip 1pt 72.0 & \hskip 2pt 49.4 & \hskip 2pt 50.9 \hskip 10pt \\
$\lambda=0.5$  & \hskip -10pt 51.5  & \hskip 1pt 70.9 & \hskip 2pt 46.9 & \hskip 2pt 50.0 \hskip 10pt \\
$\lambda=0.75$  & \hskip -10pt 59.5  & \hskip 1pt 69.9 & \hskip 2pt 43.9 & \hskip 2pt 48.5 \hskip 10pt \\
$\lambda=1$  & \hskip -10pt 66.3  & \hskip 1pt 68.9 & \hskip 2pt 40.4 & \hskip 2pt 46.6 \hskip 10pt \\

\bottomrule
\end{tabular*}
\caption{\textit{Percentage of sequences classified as functional, average intra-dataset distance (\( D_{P^2-P^2} \)), and average minimum distance from the positively reintegrated dataset (\( D_{P^2-\mathcal{D}_T^+} \)) for samples coming from reintegrated models trained for Chorismate Mutases at different values of the reintegration strength \( \lambda \).}}
\label{table4}
\end{table}
\newpage{}
\section{Group I Intron Rybozimes}
\subsection{Reintegration $\mathcal{D}_T$ Dataset}

\begin{figure}[H]
\begin{center}
\includegraphics[width=0.6\columnwidth]{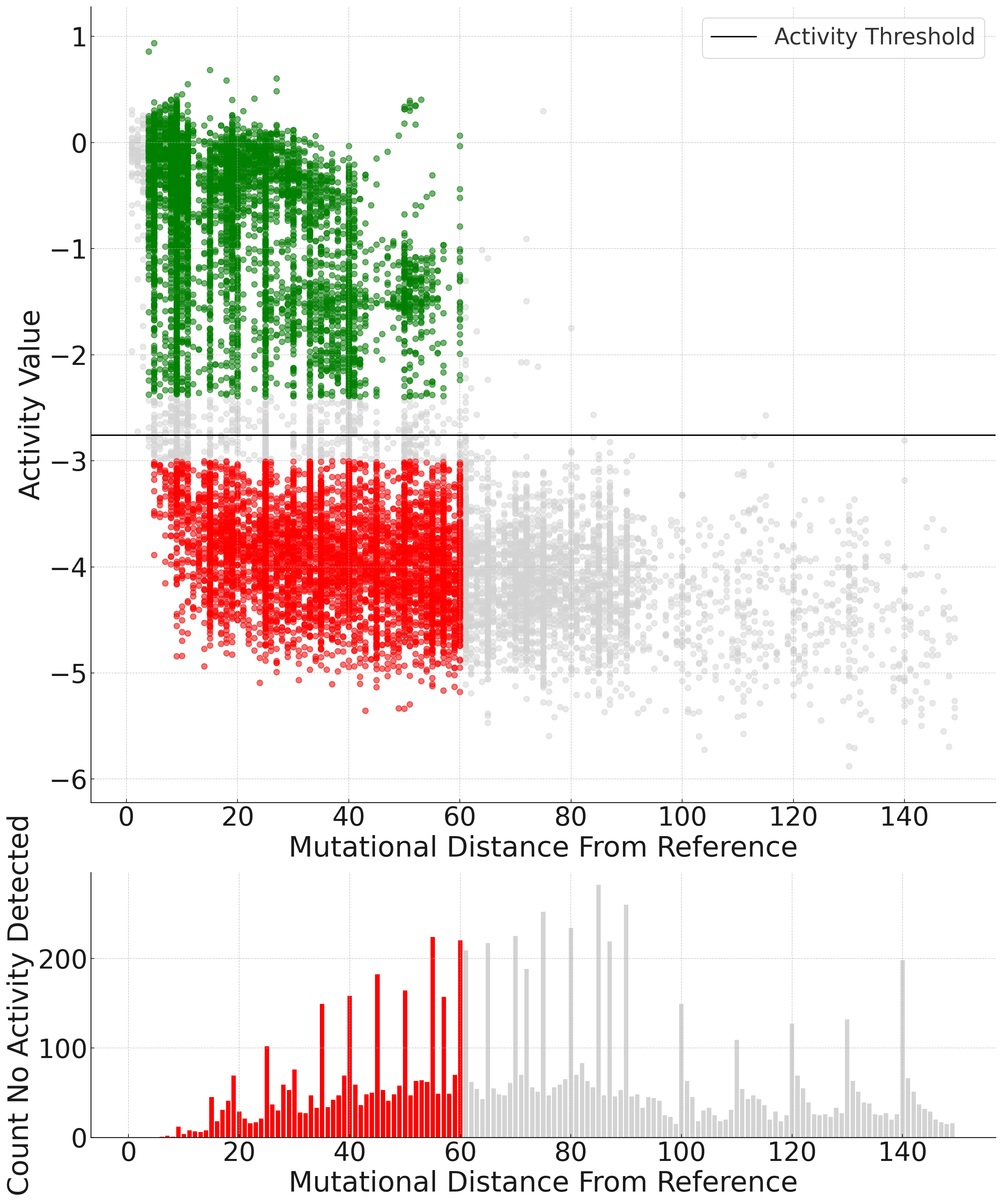}
\end{center}
\caption{\textit{Self-splicing experimentally annotated dataset from \cite{Lambert2024-rd} The top plot shows the activity versus distance from the reference sequence for this dataset. The horizontal line represents the activity threshold (-2.76) set in \cite{Lambert2024-rd}, where sequences with activity above the threshold are considered active, while those below are considered inactive. Colored sequences (green/red) represent the sequences used in the $\mathcal{D}_T$ reintegration dataset. Green dots represent reintegrated positive examples ($\mathcal{D}_T^+$), and red dots represent reintegrated negative examples ($\mathcal{D}_T^-$). Sequences with activity values close to the threshold ($-3.00 < x < -2.40$) were excluded from reintegration to avoid ambiguous signals. The bottom plot shows the number of sequences with an activity of $-\infty$ for each bin of distance.}}
\label{fig:DTazo}
\end{figure}

The ribozyme reintegration dataset $\mathcal{D}_T$ consists of 14099 experimentally annotated sequences out of the 24071 \cite{Lambert2024-rd} tested ones, all of length 197. Fig.~\ref{fig:DTazo} illustrates the experimentally determined activity and mutational distance from the Azoarcus reference sequence for all tested sequences in \cite{Lambert2024-rd}. Positively reintegrated sequences are shown in green, while negatively reintegrated sequences are shown in red. Reintegrated sequences have between 4 and 60 mutations from wildtype, beyond which no reliable activity signal was detected. We also excluded sequences in the direct vicinity of the activity threshold from the reintegration set, to avoid noisy activity annotations to be reintegrated. 

\subsection{Training Procedure and choice of $\lambda$ }

For the training of the two reintegrated models on the RNA group I intron self-splicing ribozymes, we employed a previous version of the adabmDCA algorithm, implemented in JAX. To ensure convergence, we set any negative effective frequencies to zero, consistent with our approach in all previous cases. The training procedure followed the standard Pytorch adabmDCA protocol, utilizing 10,000 Monte Carlo chains and 10 sweeps per gradient update, but with a reduced learning rate of 0.01. Training was terminated once the Pearson correlation between the empirical and training correlation matrices, \( C_{ij}^{\text{empirical}}(a,b) \) and \( C_{ij}^{\text{training}}(a,b) \), reached a threshold of 0.95. 
It is important to note that the non-reintegrated original DCA model, was trained using the eaDCA algorithm. However, in this case, the eaDCA procedure failed to converge within \( 10^4 \) training steps, necessitating the use of the adabmDCA approach. \\
The choice of $\lambda$ for the two reintegrated models is not straightforward for two main reasons. First, we are measuring actual experimental activity, which means we cannot run multiple experiments to tune the parameter. Second, we used a slightly more sophisticated adjustment function. \\
For the REINT case, $\lambda$ was set to 5000. The number of sequences in $\mathcal{D}_T^+$ is 5455, and the number of sequences in $\mathcal{D}_T^-$ is 8644. Accordingly, the weight assigned to positively reintegrated sequences is $w = 1/5455$, while for negatively reintegrated sequences, it is $w = 1/8644$. The value of $\lambda = 5000$ was chosen such that $w \cdot \lambda \approx 1$, a setting that consistently produced good results in computational examples. \\
In the case of the REINT BS0 model, the average size of the 14 $\mathcal{D}_T^+$ bins was 390 sequences, while the 14 $\mathcal{D}_T^-$ bins averaged 620 sequences. Ideally, a $\lambda$ value of around 400–500 should have been selected for the same reason as before, however, convergence issues arose during training. To address this, we selected the largest $\lambda$ value that led to convergence within 3 hours (with a learning rate of 0.01), resulting in $\lambda = 100$.  

\subsection{Experimental validation of activity of predicted RNA sequences}

\begin{figure}[H]
\begin{center}
\includegraphics[width=\columnwidth]{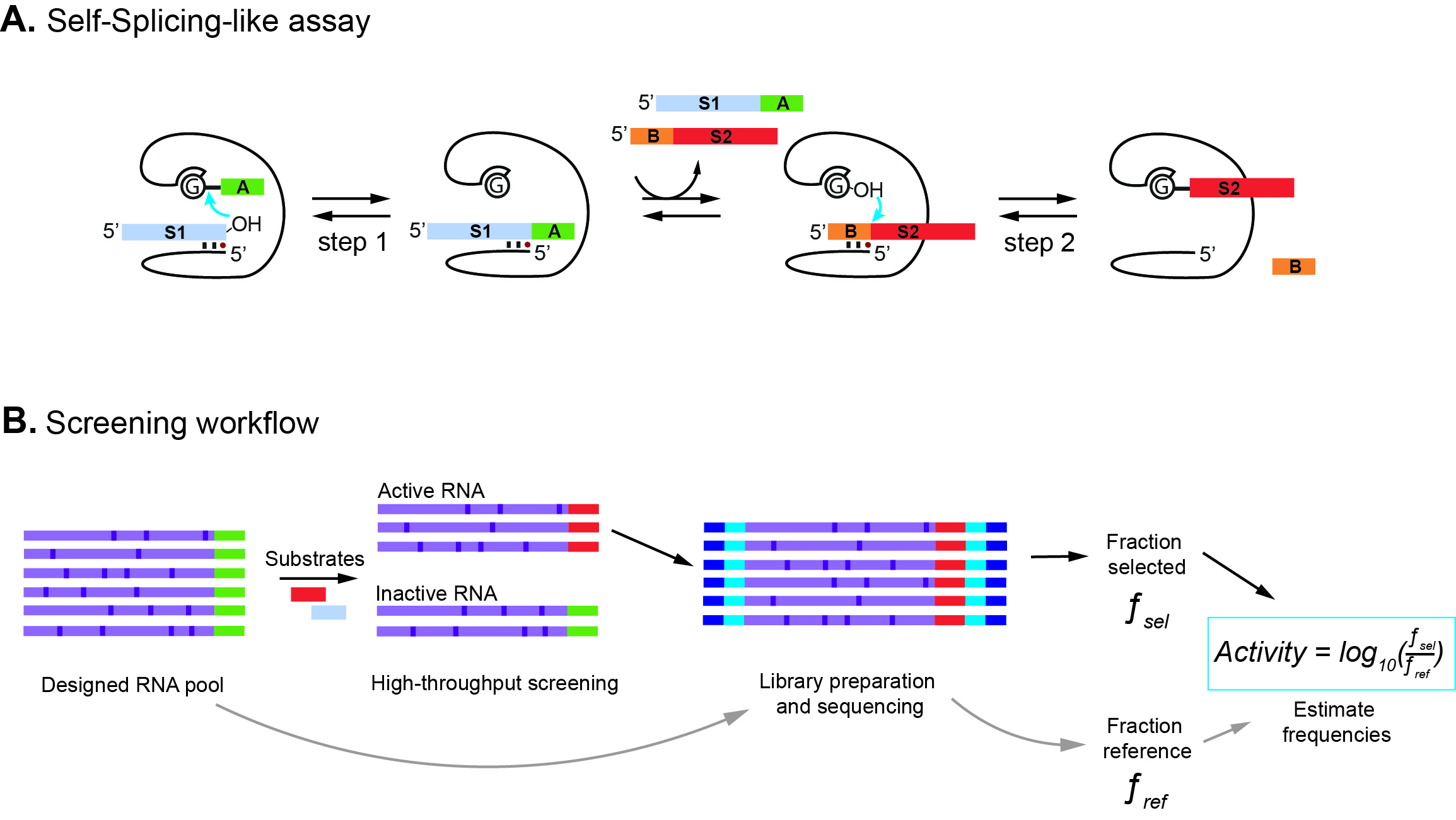}
\end{center}
\caption{\textit{High-throughput screening of functional RNA. (A) Self-splicing-like assay. (B) Scheme of the high-throughput screening of functional RNA sequence. Adapted from \cite{Lambert2024-rd}.}}
\label{HT_RNA}
\end{figure}

The goal of the experimental procedure was to identify from the designed variants the active one. The selection procedure was based on the self-splicing-like assay (Fig. \ref{HT_RNA}A) \cite{Lambert2024-rd}. Initially, exon sequence A is covalently attached to the 3'-end of the ribozyme molecule. During the first step of reaction, ribozyme binds to "substrate S1" and through recombination events forms the covalent link between "substrate S1" and A RNA fragments. Then the substrate detaches the formed "complex S1-A" and binds to the new "substrate B-S2". After, the ribozyme covalently attaches the "S2" part of the substrate to the 3'-end of itself. 
The variants that have attached the "S2" part can be specifically selected during the library preparation step. The frequency of each RNA variant in the initial pool was used as the reference value and taken into account in the final calculation of the activity of each variant (Fig. \ref{HT_RNA}B). This method was successfully applied to evaluate the activity of designed variants of the \textit{Azoarcus} ribozyme \cite{Lambert2024-rd}. 

\subsubsection{RNA production}

Designed RNA sequences were ordered as corresponding ssDNA templates with the exon sequences at the 3'-end ('AATCCGTTGGTGCTG'), and the T7 promoter at the 5'-end ('TAATACGACTCACTATA'). The ssDNA pool contained 12000 sequences was purchased from Twist Bioscience. The primers and RNA substrates were ordered from the Integrated DNA Technologies (IDT). 

All reactions were performed using RNA DNAse/RNAse-Free Water (UltraPureTM Distilled Water, Invitrogen). The DNA pool was amplified by PCR (16 cycles) using the KAPA HiFi HotStart ReadyMix (Roche) and primers F\_PCR and R\_PCR, after which the samples were purified using the NucleoSpin Gel and PCR Cleanup kit (Macherey-Nagel). The RNA pool was transcribed from the amplified DNA pool using the HiScribe T7 High Yield RNA Synthesis Kit (New England Biolabs, NEB) for 4h at 37°C in a dry bath (MyBlockTM mini dry bath). Afterwards, DNAse I treatment was performed with 10U of DNAse I in 1X DNAse I Buffer (NEB), samples were incubated for 10min at 37°C in a dry bath. 

An equal amount of phenol-chloroform (Ambion) was added to the reaction. Samples were vortexed for 1min and centrifuged for 4min at 11000rpm in a MiniSpin centrifuge (Eppendorf). The upper phase was transferred in 0.1 volume of 3M Sodium acetate (Sigma) with subsequent addition of 2.5 volumes of cold 100\%. RNA samples were precipitated overnight at -20°C. 

The sample was centrifuged for 1h at 14rpm at 4°C (Centrifuge 5418 R, Eppendorf). After, all supernatant was removed and the pellet was gently washed with 150$\mu$l of 70\% cold ethanol twice. The left ethanol was evaporated using a vacuum concentrator (Concentrator Plus, Eppendorf). The dry pellets were resuspended in 40$\mu$l of water. After 50$\mu$l of loading dye (90\% formamide, 100 mM EDTA (éthylènediaminetétraacétique), 0.1\% xylene cyanol, 0.1\% bromophenol blue) was added to each sample.

The polyacrylamide gel (20cm $\times$ 20cm) with 8M urea was prepared using the ROTIPHORESE DNA sequencing system (Carl Roth). The samples were loaded onto the 8\% urea PAGE. The gel was run for 1h at 420V. 
The gel covered by transparent plastic film was placed on Thin Layer Chromatography sheets topped with silica gel (Macherey-Nagel) and illuminated by a UV lamp at 254nm. The sections of the gel corresponding to the produced RNA were then cut with a sterile scalpel and transferred to a new 1.5ml Eppendorf tube. The gel pieces were crushed using a 1ml pipette tip. To each sample 500~$\mu$l of 0.3M sodium acetate was added. The tubes were incubated at 26°C for 5h at 450rpm in a ThermoMixer Dry Block (ThermoMixer F1.5, Eppendorf). After incubation, the upper phase was transferred to columns with a 0.22$\mu$m filter (Corstar) and centrifuged for 4 minutes at 11,000rpm. Then, 2.5 volumes of cold 100\% ethanol were added to each solution. Samples were left at -20°C overnight.

The sample was centrifuged for 1h at 14 rpm at 4°C. The supernatant was removed and the pellet was washed two times with 150 $\mu$l of cold 70\%. The residual ethanol was evaporated using a vacuum concentrator. The dry pellets were resuspended in 20$\mu$l of water. The final concentration was measured with the spectrophotometer NanoDrop One (Thermo Scientific). 

\subsubsection{Self-splicing-like assay}

The self-splicing-like reaction was perfomed according to the following protocol, 1$\mu$M of RNA pool were incubated with 25$\mu$M of "substrates S1" and "B-S2" in a reaction buffer (30mM EPPS pH7.5, 60mM $MgCl_2$) at 37°C for 1h in a final volume of 14$\mu$l. The reaction was quenched by adding EDTA to final concentration of 60mM, and cleaned using the Monarch RNA cleanup kit (New England Biolabs) with an adjusted volume of 100\%ethanol (75$\mu$l ) and binding buffer (75$\mu$l). The sample were eluted in 12$\mu$l of water. 

A control experiment without substrate addition was conducted to correct for biases in the relative quantity of each synthesized ribozyme within the corresponding pool. The initial RNA pool was diluted to 1$\mu$M in a reaction buffer with a final volume of 14$\mu$l. Without incubation, the reaction was directly quenched by adding EDTA to a final concentration of 60mM. Samples were purified using the Monarch RNA Clean up Kit following the same protocol as for the self-splicing-like assay. 

\begin{table}[H]
\begin{tabular*}{\columnwidth}{@{\extracolsep{\fill}}llllll@{}}
\toprule
Name & Sequence &  Type \\
\midrule
S1   & rCrGrCrGrArArUrUrArArCrGrCrGrArCrArArCrArU  & RNA\\
B-S2 & rGrGrCrArUrArArCrUrUrCrArArArUrArUrCrUrUrCrGrGrArArCrUrCrA  & RNA\\
F\_PCR & TAATACGACTCACTATAGTG & DNA\\
R\_PCR & CAGCACCAACGGATTCC & DNA\\
RT\_S2 & TGAGTTCCGAAGATATTTGAAGTTCC & DNA\\

\bottomrule
\end{tabular*}
\label{table:RNA_DNA_seq}
\caption{\textit{List of RNA and DNA oligos}}
\end{table}

\subsubsection{Library preparation and sequencing}
The RNA samples were prepared for sequencing using NEBNext Ultra II Directional RNA Library Prep Kit for Illumina (NEB). At the first reverse transcription step, the primer RT\_S2, which is complementary to the S2 part of the substrate, was added to the RNA sample that had undergone a self-splicing-like reaction. Similarly, for the control reaction, the primer R\_PCR, which is complementary to the exon part A, was used. During PCR amplification step each sample was barcoded using NEBNext Multiplex Oligos for Illumina (Dual Index Primers Set 1, NEB). 

The final library sequenced on a NovaSeq SP flow cell (2x250 nts, 2x800 M reads) in paired ends and with 20\% of PhiX by the NGS platform at Institut du Cerveau et de la Moelle épinière (ICM, Paris) or Institut Curie (Paris).

\subsubsection{Experimental Activity Scores from Sequencing Data}

To estimate the experimental activities from sequencing data, we followed the exact same procedure as in \cite{Lambert2024-rd}. We computed the frequencies of designed sequences under two conditions: 
\begin{enumerate}
    \item the reference condition, prior to the catalytic reaction,
    \item the reacted condition, where the substrate was mixed with the designs and incubated. 
\end{enumerate}

For both conditions, we mapped each paired-end read to the closest designed sequence using \textbf{BLAST} (version 2.12) \cite{blast}. Reads were retained only if they covered at least 70\% of the mapped designed sequence with full identity. 

We first computed the frequencies \( f_{\text{ref}} \) of designs before the catalysis, which allowed us to quantify the bias in the initial synthesized pool of RNA molecules. Next, we computed the frequencies \( f_{\text{sel}} \) of designs in the reacted sample. To determine \( f_{\text{sel}} \), we counted reads where the substrate was attached immediately after the 3' end of the design, indicating successful excision of the exon and subsequent ligation of the substrate. 

The experimental activity was then calculated as:
\[
\text{act} = \log_{10}(f_{\text{sel}}/f_{\text{ref}}) 
\]
Analysis of reverse reads was sufficient for computing the activity score. Designs with fewer than 5 reads in the pre-catalysis pool were excluded from the analysis to avoid ambiguity. Sequences with \( f_{\text{ref}} > 0 \) and \( f_{\text{sel}} = 0 \) were classified as inactive.

\subsubsection{Comparability with Previous Experiments}

Since we compare the results of our experiment against those of previous experiments \cite{Lambert2024-rd}, even if the experimental assay is identical, it is essential to address the issue of comparability of the results. Fortunately, the experimental assay presented is highly reproducible and the results of different experimental pools can be easily compared.  

The experimental data used for our reintegration in \cite{Lambert2024-rd} is already the outcome of three independent assay pools. In \cite{Lambert2024-rd}, they used 355 overlapping sequences tested in these pools to align the resulting experimental values. The measured experimental activity \(\log_{10}(f_{\text{sel}}/f_{\text{ref}})\) exhibits a correlation of up to 99\% across different pools, and aligning the results only requires the introdcution of an additive constant. Specifically, if \(\text{act}_i\) represents the activity measured in the \(i\)-th pool, then:  
\[
\text{act}_i = \log_{10}(f_{\text{sel}}/f_{\text{ref}}) + \alpha_i,
\]  
where \(\alpha_i\) is the constant used to align the values.  

In our case, since alignment with the \(P^1\) sequences is crucial, we used the same 355 overlapping sequences to align the activity values with those from the \(P^1\) pool in \cite{Lambert2024-rd}. The results of this alignment are shown in Figure~4. The comparison reveals an almost perfect match, with a linear regression slope \(m = 1.02\) and intercept \(q = 0\) ($R^2 = 0.96$). Thus, the chosen \(\alpha_i\) was set to the same value as the \(P^1\) pool from \cite{Lambert2024-rd} ($\alpha_i = -0.5103$), ensuring that the activity values are directly comparable to those presented in \cite{Lambert2024-rd}.

\begin{figure}[H]
\begin{center}
\includegraphics[width=0.7\columnwidth]{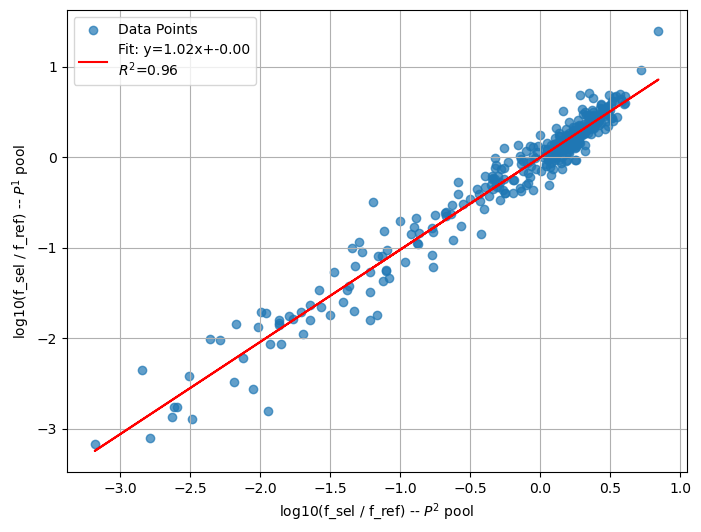}
\end{center}
\caption{\textit{Comparison of  $\log_{10}(f_{sel}/f_{ref})$ between the $P^1$ pool \cite{Lambert2024-rd} (y-axis) and $P^2$ pool (x-axis). The linear regression analysis reveals a near-perfect match with a slope of $1.02$, an intercept of $0$, with an $R^2$ value of $0.96$.}}
\label{Correlations Global}
\end{figure}

\subsubsection{Group I Intron: Tables and Violin Plot of key metrics}
\begin{figure}[H]
\begin{center}
\includegraphics[width=0.8\columnwidth, height=\textheight]{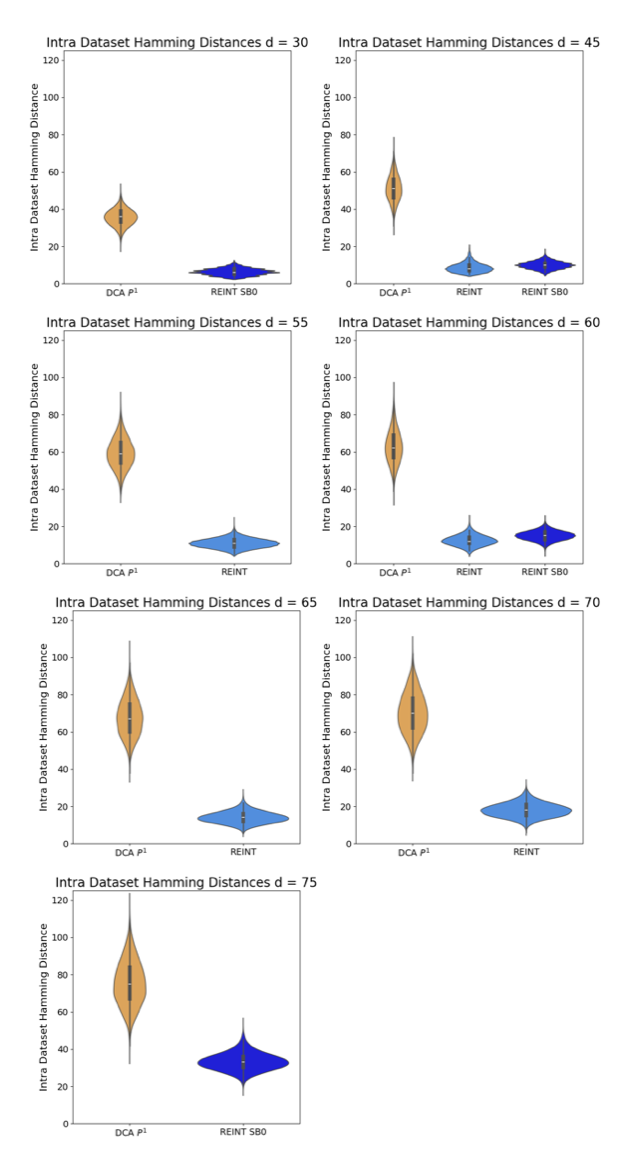}
\end{center}
\caption{\textit{Violin Plot of intra-dataset distance (\( D_{P^2-P^2} \)) for mutational distances 30, 45, 55, 60, 65, 70, and 75 from the reference sequence.}}
\label{violin_intra_distance}
\end{figure}

\begin{figure}[H]
\begin{center}
\includegraphics[width=0.8\columnwidth, height=\textheight]{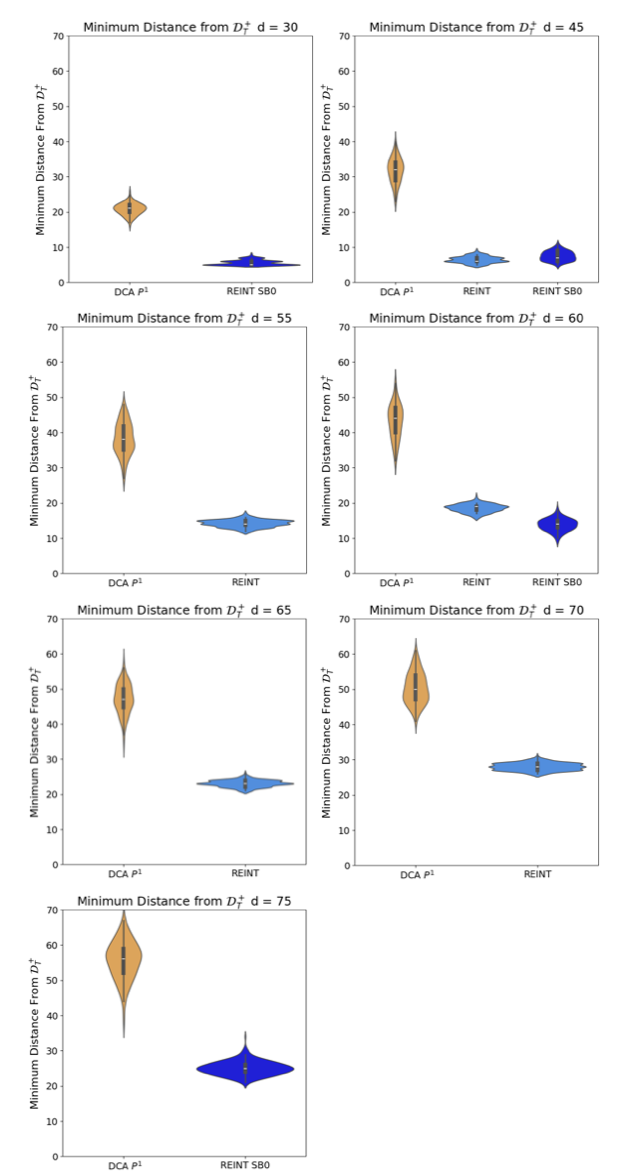}
\end{center}
\caption{\textit{Violin Plot of minimum distance from the positively reintegrated dataset (\( D_{P^2-\mathcal{D}_T^+} \)) for mutational distances 30, 45, 55, 60, 65, 70, and 75 from the reference sequence.}}
\label{violin_minimum_distance}
\end{figure}

Table \ref{table:Intra_azo} summarizes key metrics, including active fractions of sequences at different distances. To better visualize these metrics, violin plots in Figures \ref{violin_intra_distance} and \ref{violin_minimum_distance} show, respectively, the distribution of intra-dataset distances and minimum distances to the reintegrated dataset.
\begin{table}[H]
\centering
\begin{tabular*}{\columnwidth}{@{\extracolsep{\fill}}llll}
\toprule
Model (Distance) & Active \% & \( D_{P^2-P^2} \) & \( D_{P^2-\mathcal{D}_T^+} \) \\
\midrule
DCA \( P^1 \) (30)   &  43.3 & 35.5 & 20.8 \\
REINT BS0 (30)       & 99.0 & 6.4 & 5.6 \\
\midrule
DCA \( P^1 \) (45)   &  6.7 & 51.3 & 31.8 \\
REINT (45)           & 63.7 & 8.7  & 6.5 \\
REINT BS0 (45)       & 52.0 & 10.0 & 7.6 \\
\midrule
DCA \( P^1 \) (55)   &  2.0 & 59.4 & 38.3 \\
REINT (55)           & 14.4 & 11.0 & 14.2 \\
\midrule
DCA \( P^1 \) (60)   &  2.0 & 62.8 & 43.1 \\
REINT (60)           &  3.3 & 12.5 & 18.6 \\
REINT BS0 (60)       & 23.6 & 15.1 & 13.8 \\
\midrule
DCA \( P^1 \) (65)   &  0.0 & 67.7 & 47.2 \\
REINT (65)           &  2.8 & 14.2 & 23.1 \\
\midrule
DCA \( P^1 \) (70)   &  0.0 & 70.4 & 50.3 \\
REINT (70)           &  0.6 & 18.1 & 28.0 \\
\midrule
DCA \( P^1 \) (75)   &  0.0 & 75.5 & 55.7 \\
REINT BS0 (75)       &  0.7 & 33.3 & 25.1 \\
\bottomrule
\end{tabular*}
\caption{\textit{Active fraction, average intra-dataset distance (\( D_{P^2-P^2} \)), and average minimum distance from the positively reintegrated dataset (\( D_{P^2-\mathcal{D}_T^+} \)) for mutational distances 30, 45, 55, 60, 65, 70, and 75 from the reference sequence.}}
\label{table:Intra_azo}
\end{table}

\bibliographystyle{ieeetr}
\bibliography{references.bib}{}
